\DeclareSymbolFont{AMSb}{U}{msb}{m}{n}
\documentclass[reqno,centertags,11pt,a4paper,noamsfonts]{amsart}
\pdfoutput=1
\usepackage[svgnames]{xcolor}
\usepackage{graphicx}
\usepackage[english]{babel}
\usepackage[babel=true,expansion=alltext,protrusion=alltext-nott,final]{microtype}
\usepackage[margin={3cm},marginparwidth={2.5cm}]{geometry}
\usepackage[utf8]{inputenc}
\usepackage{textcomp}
\usepackage[sb]{libertine}
\usepackage[libertine,bigdelims,vvarbb]{newtxmath}
\useosf
\usepackage[varqu,varl]{inconsolata}
\usepackage[supsfam=LibertinusSerif-Sup,%
       supscaled=1.2,%
       raised=-.13em]{superiors}
\usepackage{mathtools}
\usepackage[T1]{fontenc}
\usepackage{textcomp}
\usepackage{amsthm}
\usepackage[inline]{enumitem}
\numberwithin{equation}{section}
\usepackage[normalem]{ulem}

\usepackage{float}  
\usepackage{caption}
\usepackage{subcaption} 
\usepackage{xspace}         
\usepackage[scientific-notation=true]{siunitx}      
\usepackage{pstricks}
\usepackage{tikz}
\usetikzlibrary{positioning} 
\usetikzlibrary{calc}
\usepackage{pgfplots}
\pgfplotsset{width=10cm,compat=1.9}
\usepackage{bm}
\usepackage{sidecap}
\usepackage{graphicx,color}

\usepackage{amsmath}
\DeclareFontFamily{U}{mathx}{}
\DeclareFontShape{U}{mathx}{m}{n}{<-> mathx10}{}
\DeclareSymbolFont{mathx}{U}{mathx}{m}{n}
\DeclareMathAccent{\widehat}{0}{mathx}{"70}
\DeclareMathAccent{\widecheck}{0}{mathx}{"71}

\providecommand{\mr}[1]{\href{http://www.ams.org/mathscinet-getitem?mr=#1}{MR~#1}}
\providecommand{\zbl}[1]{\href{https://zbmath.org/?q=an:#1}{Zbl~#1}}

\providecommand{\doi}[2]{\href{https://doi.org/#1}{#2}}
\newcommand{\RR}{\mathbb{R}}

\newcommand{\NN}{\mathbb{N}}

\definecolor{light_gray}{gray}{0.75}
\definecolor{lighter_gray}{gray}{0.5}
\colorlet{light_blue}{blue!20}
\definecolor{dark_green}{rgb}{0.0, 0.6, 0.0}
\definecolor{royal_blue}{rgb}{0.0, 0.22, 0.66}
\definecolor{salmon}{rgb}{1.0, 0.55, 0.41}
\definecolor{gold}{rgb}{0.8, 0.63, 0.21}
\definecolor{navy_blue}{rgb}{0.0, 0.0, 0.5}
\definecolor{crimson}{rgb}{0.79, 0.0, 0.09}
\definecolor{amethyst}{rgb}{0.6, 0.4, 0.8}
\definecolor{alizarin}{rgb}{0.82, 0.1, 0.26}
\definecolor{amaranth}{rgb}{0.9, 0.17, 0.31}
\definecolor{azure}{rgb}{0.0, 0.5, 1.0}
\definecolor{canaryyellow}{rgb}{0.82, 0.41, 0.12}
\definecolor{carrotorange}{rgb}{0.8, 0.33, 0.0}
\definecolor{cadmiumgreen}{rgb}{0.0, 0.42, 0.24}
\definecolor{copper}{rgb}{0.72, 0.45, 0.2}
\definecolor{aqua}{rgb}{0.5, 1.0, 0.83}
\definecolor{awesome}{rgb}{1.0, 0.13, 0.32}
\definecolor{candyapplered}{rgb}{1.0, 0.03, 0.0}
\definecolor{caribbeangreen}{rgb}{0.0, 0.8, 0.6}
\definecolor{indigo}{rgb}{0.0, 0.25, 0.42}

\DeclareMathOperator{\weaklystar}{\rightharpoonup\kern-2.2ex ^* \, \,}

\def\XXint#1#2#3{{\setbox0=\hbox{$#1{#2#3}{\int}$ }
\vcenter{\hbox{$#2#3$ }}\kern-.6\wd0}}

\newcommand{\R}{\mathbb R}
\newcommand{\N}{\mathbb N}

\newcommand{\ra}{\rightarrow}

\renewcommand{\phi}{\varphi}

\theoremstyle{plain}
\newtheorem{theorem}{Theorem}[section]
\newtheorem{proposition}[theorem]{Proposition}
\newtheorem{conjecture}[theorem]{Conjecture}

\newtheorem{lemma}[theorem]{Lemma}

\newtheorem*{theorem*}{Theorem}

\theoremstyle{definition}

\newtheorem{remark}[theorem]{Remark}
\newtheorem*{remark*}{Remark}

\usepackage[pagebackref=false]{hyperref}
\hypersetup{
	linktoc=all,
	bookmarksdepth=3,
	colorlinks=true,
	linkcolor=Green,citecolor=Orange,urlcolor=DarkBlue,
	pdftitle={Spectral estimations for the shifted Coulomb Hamiltonian},
	pdfauthor={Thiago Carvalho Corso, Timo Weidl, Zhuoyao Zeng}, 
	pdfsubject={},
	pdfkeywords={}} 
\begin{document}
\numberwithin{table}{section}
\title{Lieb-Thirring inequalities for the shifted Coulomb Hamiltonian}

\author[T.~Carvalho~Corso]{Thiago Carvalho Corso}
\address[T.~Carvalho Corso]{Institute of Applied Analysis and Numerical Simulation, University of Stuttgart, Pfaffenwaldring 57, 70569 Stuttgart, Germany}
\email{thiago.carvalho-corso@mathematik.uni-stuttgart.de}

\author[T.~Weidl]{Timo Weidl}
\address[T.~Weidl]{Institute of Analysis, Dynamics and Modeling, University of Stuttgart, Pfaffenwaldring 57, 70569 Stuttgart, Germany}
\email{timo.weidl@mathematik.uni-stuttgart.de}

\author[Z.~Zeng]{Zhuoyao Zeng}
\address[Z.~Zeng]{Institute of Applied Analysis and Numerical Simulation, University of Stuttgart, Pfaffenwaldring 57, 70569 Stuttgart, Germany}
\email{zhuoyao.zeng@mathematik.uni-stuttgart.de}


\date{\today}
\thanks{\emph{Funding information}:  DFG -- Project-ID 442047500 -- SFB 1481.  \\[1ex]
\textcopyright 2024 by the authors. Faithful reproduction of this article, in its entirety, by any means is permitted for noncommercial purposes.}
\begin{abstract}
In this paper we prove sharp Lieb-Thirring (LT) inequalities for the family of shifted Coulomb Hamiltonians. 
More precisely, we prove the classical LT inequalities with the semi-classical constant for this family of operators in any dimension $d\geq 3$ and any $\gamma \geq 1$. 
We also prove that the semi-classical constant is never optimal for the Cwikel-Lieb-Rozenblum (CLR) inequalities for this family of operators in any dimension. 
In this case, we characterize the optimal constant as the minimum of a finite set and provide an asymptotic expansion as the dimension grows. 
Using the same method to prove the CLR inequalities for Coulomb, we obtain more information about the conjectured optimal constant in the CLR inequality for arbitrary potentials. 
\end{abstract}
\maketitle
\tableofcontents
\setcounter{secnumdepth}{3}
\section{Introduction and statement of the main results}\label{sec:intro}
Let $V:\RR^d \ra \RR$ be a potential, such that the Schrödinger operator $-\Delta-V$ is lower semi-bounded with a
compact negative part.
We denote the negative eigenvalues of $-\Delta-V$ in increasing order by 
$$
\lambda_0<\lambda_1<\lambda_2<\cdots <0
$$
and the corresponding finite multiplicity of each $\lambda_j$ by $\mu_j\in\NN$. 
In the study of this negative spectrum the 
so-called Lieb-Thirring inequalities play an important role. 
Assume that $\gamma\geq  1/2$ for $d=1$, 
$\gamma>0$ for $d=2$ or $\gamma\geq  0$ for $d\geq  3$.
Then for any such admissible pair of $d$ and $\gamma$ 
   there exists a finite constant $R  \in \RR_+$, 
   such that for all $V\in \mathrm{L}^1_\mathrm{loc}(\RR^d)$ with $V_+\in \mathrm{L}^{\gamma+d/2}(\RR^d)$ it holds
   \footnote{For the positive and negative parts of real numbers or
   self-adjoint operators we write $a_\pm\coloneqq(|a|\pm a)/2$.}
    \begin{align}
    \mathrm{Tr}(-\Delta-V)_-^\gamma 
    \leq  
    R  \, L_{\gamma,d}^\mathrm{cl} \int_{\RR^d} V_+ ^{\gamma+d/2} \mathrm{d}x.
     \label{eq:LT}
     \end{align}
    Here  $L_{\gamma,d}^\mathrm{cl}$ stands for the semi-classical constant
    \begin{align}
    L_{\gamma,d}^\mathrm{cl} \coloneqq  \frac{\Gamma(\gamma+1)}{(4\pi)^{d/2} \, \Gamma(\gamma+1+d/2)}\, \label{eq:semiclassicalconstant}
    \end{align}  
with $\Gamma$ denoting the standard gamma function.

If $\gamma>0$, the expression $\mathrm{Tr}(-\Delta-V)_-^\gamma = \sum_{j } \mu_j|\lambda_j|^\gamma$ is often called 
the Riesz mean of order $\gamma$
of the negative eigenvalues, taking into account their 
multiplicities. On the other hand,
if $\gamma=0$, the left hand side
$\mathrm{Tr}(-\Delta-V)_-^0 = \sum_{j } \mu_j$ equals the
total multiplicity of all negative eigenvalues. In this case
\eqref{eq:LT}
is commonly known as Cwikel-Lieb-Rozenblum (CLR) inequality. 

The CLR-inequality was proven independently by Cwikel \cite{Cwi77}, Lieb \cite{Lie80} and Rozenblum \cite{Roz76}. 
Lieb and Thirring \cite{LT76} proved the cases $\gamma>1/2$ in $d=1$ and $\gamma>0$ in $d \geq  2$. 
The "limit" case $\gamma=1/2$ in $d=1$ was solved by Weidl \cite{Wei96}. Note that \eqref{eq:LT} fails for $\gamma<1/2$
if $d=1$ and for $\gamma=0$ if $d=2$.

As the validity of the bound \eqref{eq:LT} has been completely settled, nowadays research focuses 
on the optimal values $R_{\gamma,d}$ of the constants $R$.  
For the best known bounds on the optimal constant up to date we refer the reader to the book \cite{FLW23} and to the recent works \cite{CCR24,Cor24}. 
In particular, semi-classical analysis shows that $R_{\gamma,d}\geq  1$.

In this paper we are interested in the family of CLR and LT inequalities for a special kind of Schrödinger operators, namely the shifted Coulomb Hamiltonian in $\mathrm{L}^2(\RR^d)$, defined as 
\begin{align*}
    -\Delta - \frac{\kappa}{|x|} + \Lambda \quad\quad  \mbox{for $\kappa, \Lambda >0$}.
\end{align*}
This operator stands out for two reasons. For one thing, its spectrum can be computed explicitly, which allows for a direct analysis of
the resulting expressions. 
For another, it is one of the physically most relevant Schr\"odinger operators, as it serves as a basic quantum model for non-interacting electrons bound to a point nuclei with charge $\kappa>0$. 
Despite these facts, there are (to the best of our knowledge) only few works attempting to explicitly compute optimal constants in 
CLR and LT inequalities restricted to this family of operators, 
namely \cite{FLW23} and \cite{Sel24}.

The goal of this work is to fill in this gap. More precisely, our main contributions here are the following:
\begin{enumerate}[label=(\roman*)]
\item We prove that for the family of shifted Coulomb potentials
$V=\kappa|x|^{-1}-\Lambda$ the LT inequality \eqref{eq:LT} holds true with $R=1$ for all dimensions $d\geq  3$ if $\gamma \in [1,d/2)$.
\item On the other hand we prove that 
in any dimension $d\geq  3$
the CLR inequality (i.e. $\gamma=0$) restricted to the full class of shifted Coulomb Hamiltonians does not hold with $R=1$.
Moreover, we characterize the optimal constant in this case and give an asymptotic expansion as the dimension increases.
\item We prove that the conjectured value of the optimal CLR constant \emph{for arbritary potentials} \cite{GGM78}, which is given by the minimization of a specific function over integers, can be reduced to a minimization on an interval of length of order $d$ 
around the point $d^2/6$.
As a by-product of this analysis we obtain an explicit asymptotic expansion of the conjectured optimal value as the dimension $d$ grows.  
\end{enumerate}

We now present the precise statements of our main results.

\subsection{Main results (i)} 
 Our first result shows that the semi-classical constant is optimal for the Lieb-Thirring inequalities for the family of shifted Coulomb Hamiltonians with $\gamma\geq  1$. This result extends a result by Frank, Laptev, and Weidl \cite[Section 5.2.2]{FLW23} for the case $d=3$ to all dimensions $d \geq  3$.

\begin{theorem}[Optimal LT inequalities for the shifted Coulomb Hamiltonian]
\label{thm: LT Coulomb} 
    Let $d\geq  3$ and $\gamma\in[1,d/2)$. Then for any $\kappa,\Lambda>0$ we have
    \begin{align}
        \mathrm{Tr} \left( -\Delta - \frac{\kappa}{|x| } + \Lambda \right)_-^\gamma 
        < \, 
        L_{\gamma,d}^\mathrm{cl} \int_{\RR^d} \left( \frac{\kappa}{|x|} - \Lambda\right)_+ ^{\gamma+d/2} \mathrm{d} x, \label{eq:LT Coulomb opt const}
    \end{align}
    where $L^{\rm cl}_{\gamma,d}$ is the semi-classical constant defined in~\eqref{eq:semiclassicalconstant}.
\end{theorem} 

We refer to \eqref{eq:Riesz means formula} and \eqref{eq:LT r.h.s.} 
below for the explicit formulae of both sides in \eqref{eq:LT Coulomb opt const}.

Since the family of shifted Coulomb potentials is closed with
respect to the shift in energy, the general case 
$\gamma\in[1,d/2)$ follows by the Aizenman-Lieb argument \cite{AL78}
from \eqref{eq:LT Coulomb opt const} with $\gamma=1$.
\footnote{The upper restriction $\gamma<d/2$ follows naturally from the fact
that $(\kappa|x|^{-1}-\Lambda)_+\in \mathrm{L}^p(\mathbb
{R}^d)$ only for $p<d$.}
This case is of particular interest. Here one has, 
see \eqref{eq:LT r.h.s. gamma=1},
\begin{align}\label{eq:PhSpcVolgamma1}
L_{1,d}^\mathrm{cl} \int_{\RR^d} \left( \frac{\kappa}{|x|} - \Lambda\right)_+ ^{1+d/2} \mathrm{d} x = \frac{2^{2-d} \Lambda^{1-d/2} \kappa^d}{d!(d-2)}\,.
\end{align}
We point out that 
the bound \eqref{eq:LT Coulomb opt const} is strict.
For $d\geq  4$ this also follows via the Aizenman-Lieb argument
from the following somewhat stronger 
inequality in the case $\gamma=1$, which we shall actually prove, cf. \eqref{eq:LT esitmate final}.

\begin{proposition}[Improved estimate]  \label{remark:improvedestimate}
For $d\geq  4$ it holds
\begin{align}
     \mathrm{Tr} \left( -\Delta - \frac{\kappa}{|x| } + \Lambda \right)_- 
         \leq  
         \left(\frac{2^{2-d} \Lambda^{1-d/2} \kappa^d}{d!(d-2)} - \frac{\kappa^2}{4(d-1)(d-2)^2}\right)_+ \label{eq:LT strict ineq gamma1}  
\end{align}
for any $\kappa, \Lambda >0$.
\end{proposition}

For $d=3$ the bound \eqref{eq:LT strict ineq gamma1} itself does not hold. However,
straightforward computations yield 
the following modified elementary upper and lower bounds.

\begin{proposition}[Sharp corrections to Lieb-Thirring for $d=3$]
\label{thm:LTnextorder} 
For all $\kappa, \Lambda >0$ we have
\begin{align}
      &\mathrm{Tr} \left( -\Delta - \frac{\kappa}{|x| } + \Lambda \right)_- 
         \leq  
         \left(\frac{\kappa^3}{12\sqrt{\Lambda}} - \frac{\kappa^2}{8}+ 
\frac{\sqrt{\Lambda}\varkappa}{24} \right)_+,\quad \varkappa = \sqrt{\Lambda}\left(2\left\lceil\frac{\kappa}{2\sqrt{\Lambda}}
\right\rceil-1\right)\,,
          \label{eq:LT strict ineq gamma1 d=3} \\
         &\mathrm{Tr} \left( -\Delta - \frac{\kappa}{|x| } + \Lambda \right)_- \geq  \left( \frac{\kappa^3}{12\sqrt{\Lambda}} - \frac{\kappa^2}{8}- \frac{\sqrt{\Lambda} \kappa}{12} \right)_+ . \label{eq:LTsharplowerbound} 
         \end{align}  
These inequalities are sharp as equality is achieved in~\eqref{eq:LT strict ineq gamma1 d=3} whenever $ \kappa/\sqrt{\Lambda} $ is an odd natural number and in~\eqref{eq:LTsharplowerbound} whenever $\kappa/\sqrt{\Lambda} $ is an even natural number. 
\end{proposition}

Note that for $d=3$ the expression \eqref{eq:PhSpcVolgamma1}
turns into $\kappa^3/12\sqrt{\Lambda}$, 
see \eqref{eq:LT r.h.s. gamma=1 d=3}.
Since $\kappa^2/8 - \sqrt{\Lambda} \varkappa/24 \geq \kappa^2/8 - \Lambda (\kappa / \sqrt{\Lambda }+1)/24>0 $
whenever the l.h.s. of \eqref{eq:LT strict ineq gamma1 d=3} is positive, that is for $\kappa/\sqrt{\Lambda}>2$,
the bound \eqref{eq:LT Coulomb opt const} is strict for $d=3$, too.

The term $-\kappa^2/8$ in \eqref{eq:LT strict ineq gamma1 d=3} and \eqref{eq:LTsharplowerbound} is related
to the so-called Scott correction \cite{SW87}.
The bounds in Proposition~\ref{thm:LTnextorder} are instructive, since the asymptotic envelopes of the eigenvalue sum, including 
the semiclassical term, the Scott correction and the oscillatory third
term, serve also as sharp universal upper and lower bounds. For an illustration see Figure \ref{fig:LT d=3}. 

\begin{figure}[ht]
    \centering
    \begin{minipage}{0.8\textwidth}
        \centering
        \resizebox{1\linewidth}{!}{\input{plots/LT_d3.pgf}} 
        
    \end{minipage}
    \caption{Behavior of the difference between l.h.s. and r.h.s. in \eqref{eq:LT strict ineq gamma1} for $d=3$, $\Lambda=1$ and $\kappa\in (2,20]$ oscillating between the correction terms from Proposition~\ref{thm:LTnextorder}.} 
    \label{fig:LT d=3}
\end{figure}

Finally, let us point out that for $d\geq  4$ the correction term in \eqref{eq:LT strict ineq gamma1} is {\em not} related to a Scott type term, since the asymptotics of the eigenvalue sums for the Coulomb Hamiltonian show a different behavior in higher dimensions \cite{Sol96}. 
The sole purpose of this term is to show strictness
of \eqref{eq:LT Coulomb opt const}.

\subsection{Main results (ii)}
Our next result concerns the CLR inequality for the Coulomb Hamiltonian. 
In \cite{Sel24} it is claimed that the optimal constant of the CLR inequality for the shifted Coulomb Hamiltonian is given by the semi-classical constant in case of $d\geq  6$.
However, this turns out to be incorrect\footnote{For instance, for $d=6$ and $\kappa/\sqrt{\Lambda}=11.1$, we have $\mathrm{Tr}(-\Delta-\kappa/|x|+\Lambda)_-^0=121$, but $L^\mathrm{cl}_{0,d}\int (\kappa/|x|-\Lambda)_+^{d/2}\mathrm{d}x\approx 81.81$.}.
In fact, as we shall see, for any $d\geq  3$ one can find (uncountably many) $\kappa, \Lambda>0$, s.t. 
\begin{align*}
    \mathrm{Tr} \left( -\Delta - \frac{\kappa}{|x| } + \Lambda \right)_-^0 
        >
        L_{0,d}^\mathrm{cl} \int_{\RR^d} \left( \frac{\kappa}{|x|} - \Lambda \right)_+ ^{d/2} \mathrm{d} x\,,
\end{align*} 
see Figure \ref{fig:Rd vs Qd}.

Let us perform the corresponding analysis.
For $d\geq  3$ we define the function   
\begin{align}
    Q_d(t)\coloneqq \left( t+\frac{d-1}{2}\right)^{-d}\left(t+\frac{d}{2}\right)\prod_{j=1}^{d-1} (t+j) \,,\quad t\in\R\,, t\neq -\tfrac{d-1}{2}\,.
\label{eq:Qfunction}
\end{align}
One of the main points of our analysis will be to show that $Q_d(t)$ has a unique maximum for $t\in [0,+\infty)$, at which it is larger than one.  
In fact,  for sufficiently large $d$ we can localize 
the corresponding non-negative real 
argument $t$, at which the maximum is attained, in an interval of length $d-2$ 
around the point $d^2/6 - d+ 5/6$, see Lemma \ref{lemma:Property Q_d}.
The optimal value of $R$ in the CLR inequality \eqref{eq:LT} restricted to the shifted Coulomb Hamiltonian will correspond 
to the maximal value of $Q_d(t)$ over a corresponding set of 
natural numbers. 

Indeed,
set $Q_3^*\coloneqq Q_3(0)$ as well as 
\begin{align*}
    Q_d^* \coloneqq \max \left\{ Q_d(\ell) : \ell \in \N_0 \land \left\lfloor \frac{d^2}{6} - \frac{3d}{2} + \frac{7}{3} \right\rfloor \leq  \ell \leq  \left\lceil \frac{d^2}{6} - \frac{d}{2}-\frac{2}{3} 
    \right\rceil \right\}
    \quad\text{for}\quad d\geq  4\,.
\end{align*}
It turns out that $Q_d^*$ is the optimal choice for the 
factor $R$ and that
$Q_d^*>1$ for all $d\geq  3$, 
see Remark \ref{rem:Q*}. 
In particular, the optimal constant in the CLR inequality for the family of shifted Coulomb Hamiltonians is strictly larger than the semi-classical constant 
\emph{in any dimension $d \geq  3$}. 

\begin{theorem}[Optimal CLR inequalities for the shifted Coulomb Hamiltonian]\label{thm: CLR Coulomb} 
    For all $d\geq  3$ it holds that 
    \begin{align*}
         \inf \left\{ R \in \R_+ : \mathrm{Tr} \left( -\Delta - \frac{\kappa}{|x| } + \Lambda \right)_-^0 
        \leq  
        R \,   L_{0,d}^\mathrm{cl} \int_{\RR^d} \left( \frac{\kappa}{|x|} - \Lambda \right)_+ ^{d/2} \mathrm{d} x
         \text{ for all }\kappa, \Lambda>0 \right\} = Q_d^* >1.
    \end{align*}
\end{theorem}

Furthermore we can derive the following asymptotic expansion 
for $Q_d^*$ as $d\ra +\infty$.

\begin{proposition}[Asymptotic expansion in high dimensions]\label{cor:asymptotic}
It holds
\begin{align*}
    Q_d^\ast  
    = 1 + \frac{3}{2d} + \frac{45}{8 d^2} + \mathcal{O}\left({d^{-3}}\right) \quad\text{as}\quad d\ra +\infty\,. 
\end{align*}
\end{proposition}

\begin{remark}
In the process of proving
Theorem \ref{thm: CLR Coulomb} we shall see that 
that the maximal excess factor $R=Q^*_d$ 
in the CLR inequality \eqref{eq:LT} restricted to the shifted Coulomb Hamiltonian 
is achieved in a regime of $d^2/6 + \mathcal{O}(d)$ distinct eigenvalues as $d\ra +\infty$. 
Hence, Theorem~\ref{thm: CLR Coulomb} provides --- besides the one in \cite{GGM78} --- another class of potentials for which the optimal constant in the CLR inequality is strictly larger than both the semi-classical constant {\em and} the one-particle constant in high dimensions.
\end{remark}

\subsection{Main results (iii)}
Our last result concerns an hypothesis on the optimal CLR constant \emph{for arbritary potentials} proposed by Glaser, Grosse and Martin \cite{GGM78}.
We briefly recall their conjecture. 
As above, let $R_{0,d}$ be the
optimal value of the constant $R$ in \eqref{eq:LT} 
with $d\geq  3$ and $\gamma=0$ considered on all 
potentials $V\in \mathrm{L}^1_{\rm loc}(\R^d)$ with $V_+ \in \mathrm{L}^{d/2}(\R^d)$.

For $d \geq  3$, we define the function 
\begin{align}
\label{eq:ad}
    A_d(t)\coloneqq   
\left(t + \frac{d}{2}\right)^{1-\frac{d}{2}}
\left(t+\frac{d}{2}-1\right)^{-\frac{d}{2}} 
\prod_{k=1}^{d-1} (t+k) \,,\quad t\in\R\,,t\neq -\tfrac{d}{2}\,,t\neq -\tfrac{d}{2}+1\,,
\end{align}
and set $A_d^* \coloneqq \sup \left\{ A_d(\ell)  : \ell \in \N_0 \right\}$.
\begin{conjecture} [{\cite{GGM78}}]
It holds that 
\begin{align*}
    R_{0,d} = A_d^*. 
\end{align*}
\end{conjecture}

The structure of the function $A_d$ and the constant $A_d^*$ in the aforementioned conjecture are quite similar to the one of $Q_d$ and $Q_d^*$ from Theorem~\ref{thm: CLR Coulomb}. 
In fact, it is not hard to see that $Q_d \leq  A_d$. Therefore, our results in Theorem~\ref{thm: CLR Coulomb} do not contradict this conjecture. 

Adapting our analysis of $Q_d$ to $A_d$ we can show that $A_d^*$ and $Q_d^*$ behave quite similar as the dimension increases. 
The following theorem gives the precise formulation of this result.

\begin{theorem}[On the conjectured optimal CLR excess factor] \label{thm:CLRconjecture} 

For $d=3,4$ we have $A_d^* = A_d(0)$ and for $d\geq  5$ we have
\begin{align*}
    A_d^* = \max\left\{ A_d(\ell): \ell \in \N_0 \wedge 
    \left\lfloor \frac{d^2}{6}-\frac{3d}{2}+\frac{5}{3} \right\rfloor 
    \leq  \ell \leq  
    \left\lceil \frac{d^2}{6} - \frac{d}{2} - 1 \right\rceil    
    \right\}.
\end{align*}
Moreover, $A^*_d$ and $Q_d^*$ satisfy the asymptotic relations
\begin{align}
    A_d^* = Q_d^\ast + \mathcal{O}(d^{-3}) = 1 + \frac{3}{2d} + \frac{45}{8d^2} + \mathcal{O}\left({d^{-3}}\right)
    \quad\text{as}\quad d\ra +\infty\,.
    \label{eq:CLRconjecture_expansion}
\end{align}
\end{theorem}

\begin{remark}  Theorem~\ref{thm:CLRconjecture} shows that the conjectured value of the optimal constant in the CLR inequality approaches the semi-classical constant with convergence rate of 
order $1/d$ 
as the dimension increases, and that this value is 
(almost) achieved by the shifted Coulomb Hamiltonian up to an error of 
order $1/d^3$. To the best of our knowledge, both results are 
new. 
\end{remark}

\subsection{Outline of the paper}
In Section \ref{sec: LT proof} we shall first 
recall some basic facts about the spectrum of the shifted Coulomb Hamiltonian. Then we study the case $\gamma \geq  1$ and prove Theorem \ref{thm: LT Coulomb}. 
In Section \ref{sec:CLR} we study the case $\gamma=0$ and prove Theorem \ref{thm: CLR Coulomb} as well as Proposition \ref{cor:asymptotic}. 
In Section \ref{sec:CLR conj} we study the conjectured 
optimal excess factor in the CLR inequality
for general potentials and prove Theorem \ref{thm:CLRconjecture}. 

Auxiliary properties and some elementary calculations will be presented in the appendix.

\addtocontents{toc}{\protect\setcounter{tocdepth}{1}}

\section{On the sharp LT inequalities for the shifted Coulomb Hamiltonian} \label{sec: LT proof}
\subsection{Spectrum of the shifted Coulomb Hamiltonian} \label{subsec: spec of coulomb}
Let $d\geq  3$ and $\kappa>0$. The negative spectrum of the
Coulomb Hamiltonian $-\Delta-\kappa|x|^{-1}$ consists precisely
of the eigenvalues
\begin{align*}
- \frac{\kappa^2}{(2j+d-1)^2} 
\quad\text{with multiplicities}\quad
 \frac{(d-2+j)! \, (d-1+2j)}{(d-1)! \, j!}
\quad\text{for}\quad  j=0,1,2,\dots\,.
\end{align*}
For a detailed derivation of these formulae see \cite[Section 4.2.3]{FLW23}.

Applying an energy shift $\Lambda>0$ we see that $-\Delta-\kappa|x|^{-1}+\Lambda$ does not have negative spectrum at all as long as
\begin{align*}
\eta\coloneqq \kappa/\sqrt{\Lambda}\leq  d-1\,.
\end{align*}
Assume now that $\eta>d-1$, or equivalently 
\begin{align*}
\ell \coloneqq \lceil(\eta-d+1)/2\rceil -1\geq  0\,.
\end{align*}
Then the negative spectrum of the shifted Coulomb Hamiltonian is given by the eigenvalues
\begin{align*}
    \lambda_j = - \frac{\kappa^2}{(2j+d-1)^2} + \Lambda 
    \quad\text{for}\quad j=0,1,\dots,\ell\,,
\end{align*}
with the corresponding multiplicities
\begin{align*}
    \mu_j = \frac{(d-2+j)! \, (d-1+2j)}{(d-1)! \, j!} 
    = \left(
    \begin{smallmatrix}
        d-1+j \\ d-1
    \end{smallmatrix} \right)
    + \left(
    \begin{smallmatrix}
        d-2+j \\ d-1
    \end{smallmatrix} \right)\,.
\end{align*}
Using elementary properties of binomial coefficients and the \textit{Hockey-Stick identity}, namely
$\sum_{i=r}^{n} \left( \begin{smallmatrix} i \\r \end{smallmatrix} \right) 
= \left( \begin{smallmatrix} n+1 \\r+1 \end{smallmatrix} \right) $ for $n,r\in \N$ and $n \geq  r$, the total multiplicity
for the lowest $k+1$ eigenvalues $\lambda_0,\dots,\lambda_k$ can be computed as follows
\begin{align*}
N_k 
\coloneqq \sum_{j=0}^{k} \mu_j
=  
\frac{(d+2k) (d+k-1)!}{d! \, k! }
\quad \text{for} \  k=0,1,2,\dots, \ell. 
\end{align*}
In particular, we see that 
\begin{align}
\mathrm{Tr}(-\Delta-\kappa/|x|+\Lambda)^0_- = N_\ell
=
\frac{(d+2\ell) (d+\ell-1)!}{d! \, \ell! }\,.
\label{eq:Nl}
\end{align}
For $\gamma>0$ 
the Riesz means of the shifted Coulomb Hamiltonian are given by
\begin{align}
    \mathrm{Tr} \left( -\Delta - \frac{\kappa}{|x| } + \Lambda \right)_-^\gamma 
    = \sum_{j = 0}^{\ell } 
  \mu_j
    \left( \frac{\kappa^2}{(2j+d-1)^2} - \Lambda \right)^\gamma\,. 
    \label{eq:Riesz means formula}
\end{align}
For $\gamma=1$ this simplifies as follows
\begin{align}
  \label{eq:Riesz means formula gamma=1a}
    \mathrm{Tr} \left( -\Delta - \frac{\kappa}{|x| } + \Lambda \right)_-
    &= \sum_{j = 0}^{\ell } 
    \frac{\mu_j\kappa^2}{(2j+d-1)^2} - N_\ell \Lambda \\
     &= \sum_{j = 0}^{\ell }
      \frac{\kappa^2(d-2+j)!}{(d-1)! \, j! \, (d-1+2j)}-
      \frac{(d+2\ell) (d+\ell-1)!}{d! \, \ell! }\Lambda
    \,. 
    \label{eq:Riesz means formula gamma=1b}
\end{align}
In the case $d=3$ this turns into
\begin{align}
\label{eq:Riesz means gamma=1 d=3}
 \mathrm{Tr} \left( -\Delta - \frac{\kappa}{|x| } + \Lambda \right)_-
 = \frac{(\ell+1)\kappa^2}{4}-
 \frac{(\ell+1)(\ell+2)(2\ell+3)\Lambda}{6}\,.
\end{align}

\subsection{Phase space integrals} \label{subsec: phase space int}
The integral 
in the r.h.s. of \eqref{eq:LT Coulomb opt const}
can also be computed explicitly.
It is finite for $0\leq  \gamma<d/2$ and a standard variable transformation
and the properties of the beta function yield
\begin{align}
    \int_{\RR^d} \left( \frac{\kappa}{|x|} - \Lambda \right)_+ ^{\gamma+d/2} \mathrm{d} x 
    = (4\pi)^{\frac{d}{2}}\frac{\Lambda^\gamma \eta^d}{2^{d-1}}  \frac{\Gamma(\gamma+1+d/2)\Gamma(d/2-\gamma)}{\Gamma(d+1)\Gamma(d/2)}\,. \label{eq:gammarestriction}
\end{align} 
Combining \eqref{eq:gammarestriction} 
with the semi-classical constant
 \eqref{eq:semiclassicalconstant} we see that the r.h.s. 
 of \eqref{eq:LT Coulomb opt const} reads as follows
\begin{align}
    L_{\gamma,d}^\mathrm{cl} \int_{\RR^d} \left( \frac{\kappa}{|x|} - \Lambda\right)_+ ^{\gamma+d/2} \mathrm{d} x
    = \frac{\Lambda^\gamma \eta^d}{2^{d-1}} \frac{\Gamma(\gamma+1)\Gamma(d/2-\gamma)}{\Gamma(d+1)\Gamma(d/2)} \,. \label{eq:LT r.h.s.}
\end{align}
For $\gamma=1$ this gives
\begin{align}
 L_{1,d}^\mathrm{cl} \int_{\RR^d} \left( \frac{\kappa}{|x|} - \Lambda\right)_+ ^{1+d/2} \mathrm{d} x
    = \frac{\Lambda \eta^d}{2^{d-2}} \frac{1}{d!\,(d-2)}\,, \label{eq:LT r.h.s. gamma=1}
\end{align}
and in the case of $d=3$ we get
\begin{align}
 L_{1,3}^\mathrm{cl} \int_{\RR^3} \left( \frac{\kappa}{|x|} - \Lambda\right)_+ ^{5/2} \mathrm{d} x
    = \frac{\Lambda \eta^3}{12} \,. \label{eq:LT r.h.s. gamma=1 d=3}
\end{align}
On the other hand, for $\gamma=0$ and $d\geq  3$ one has
\begin{align}
    L_{0,d}^\mathrm{cl} \int_{\RR^d} \left( \frac{\kappa}{|x|} - \Lambda\right)_+ ^{d/2} \mathrm{d} x
    = \frac{ \eta^d}{2^{d-1}\,d!}  \,. \label{eq:LT r.h.s. gamma=0}
\end{align}

\subsection{Proving the optimal LT inequality for the shifted Coulomb Hamiltonian} 

\begin{proof}[Proof of Proposition~\ref{thm:LTnextorder}]
We have $d=3$. Using the notation $\eta=\kappa/\sqrt{\Lambda}>0$
and $\ell = \lceil (\eta-2)/2 \rceil-1 = \lceil \eta/2 \rceil  -2$
we write $\ell = \eta/2+\delta_\eta-2$ with $\delta_\eta \coloneqq  \lceil \eta/2 \rceil - \eta/2\in[0,1)$. For $\eta\leq  2$  the negative spectrum
of $-\Delta-\kappa|x|^{-1}+\Lambda$ is empty. For $\eta>2$
we have
$\ell\in\N_0$ and the identity \eqref{eq:Riesz means gamma=1 d=3} gives
\begin{align}
    \mathrm{Tr} \left( -\Delta-\frac{\kappa}{|x|} + \Lambda \right)_- 
    = \Lambda\left( \frac{\eta^2(\ell+1)}{4} - \frac{(\ell+1)(\ell+2)(2\ell+3)}{6} \right)
    = \Lambda \left(\frac{\eta^3}{12} - \frac{\eta^2}{8} + \varphi(\eta)
    \right) 
    \label{eq:almost_there_again}
\end{align} 
with
\begin{align*}
    \varphi(\eta)\coloneqq \left(\frac{\delta_\eta(1-\delta_\eta)}{2} -\frac{1}{12}\right)\,\eta - \frac{\delta_\eta(1-\delta_\eta)(1-2\delta_\eta)}{6} \,,
    \quad \eta>2\,.
\end{align*}
Rewriting $\eta=2\ell-2\delta_\eta+4=2m+2\varepsilon$
with $m=\ell+1\in\N$ and $\varepsilon=1-\delta_\eta\in (0,1]$, this turns into
\begin{align}
\label{eq:phi eta epsilon}
\phi(\eta)=\phi(2m+2\varepsilon)= -\frac{2}{3} \varepsilon^3+ \frac{1-2m}{2} \varepsilon^2 + m \varepsilon - \frac{m}{6}
\,.
\end{align}
Elementary computations show that for fixed $m\in\N$
and $\varepsilon\in (0,1]$ we have
\begin{align*}
-\frac{\eta}{12}=-\frac{m+\varepsilon}{6}\leq  \phi(\eta)
\leq  \frac{2m+1}{24}=\frac{2\lceil\eta/2\rceil-1}{24}\,.
\end{align*}
Equality is attained for each $m\in\N$
on the l.h.s. for $\varepsilon=1$ and in the
limit $\varepsilon\to +0$, while on the r.h.s. for $\varepsilon=1/2$.
If we put this back into \eqref{eq:almost_there_again}, 
we arrive at
\begin{align}
    \Lambda \left(\frac{\eta^3}{12} - \frac{\eta^2}{8} - \frac{\eta}{12}\right) \leq  \mathrm{Tr} \left( -\Delta-\frac{\kappa}{|x|} + \Lambda \right)_- 
    \leq  \Lambda \left(\frac{\eta^3}{12} - \frac{\eta^2}{8} + \frac{2\lceil\eta/2\rceil-1}{24}\right) \quad \mbox{for $\eta >2$.} \label{eq: hopefully_the_last}
\end{align}
It remains to observe that for $0<\eta\leq  2$ the left term is less or equal to zero and the middle term vanishes.
\end{proof}

\begin{proof}[Proof of Theorem~\ref{thm: LT Coulomb}] 
It remains to prove \eqref{eq:LT strict ineq gamma1} for $d\geq  4$ and $\eta=\kappa/\sqrt{\Lambda}>d-1$.

{\em Step 1.} In what follows it is convenient to use the following (shifted) Pochhammer symbols. Let $t\in\R$.  We define
\begin{align*}
p_0(t) \coloneqq 1
\end{align*}
and
\begin{align*}
p_{m}(t) \coloneqq \prod_{k=1}^{m} (t+k) \quad \mbox{for $m \in \N$}. 
\end{align*}
By definition we see that $p_m(-1)=0$ 
for $m\in\N$ and that
\begin{align*}
p_{m}(t)=(t+m)p_{m-1}(t)\quad\text{for all}\quad t\in\R\,,\ m\in\N\,.
\end{align*}
A key property of $p_m$, which can be directly verified from its definition, is the recursive relation 
\begin{align}
   m p_{m-1}(t)  =  p_{m}(t) - p_{m}(t-1) \quad \text{for all}
   \quad t\in\R\,,\ m\in \N\,. 
   \label{eq:exact recursion pm}
\end{align}
In particular, this dentity
allows us to explicitly evaluate the following sum 
\begin{align}
   m \sum_{j=0}^{\ell} p_{m-1}(j)  
   = \sum_{j=0}^{\ell} \left(p_{m}(j) - p_{m}(j-1)\right) 
   = p_{m}(\ell)\,,\quad m\in\N\,,\ \ell\in\N_0\,. 
   \label{eq:exactsummation}
\end{align}

{\em Step 2.} Let us use this property to evaluate the sum
in \eqref{eq:Riesz means formula gamma=1a} and \eqref{eq:Riesz means formula gamma=1b}. Note that
\begin{align*}
    \frac{(d-1)! \mu_j}{(2j+d-1)^2} = 
    \frac{(d-2+j)!}{j! \, (d-1+2j)} =
    \frac{p_{d-2}(j)}{d-1+2j} = \frac{ (d-2+j)\,p_{d-3}(j)}{d-1+2j}
    \,.
\end{align*} 
This yields
\begin{align*}
    \frac{(d-1)! \mu_j}{(2j+d-1)^2} = 
    \frac{p_{d-3}(j)}{2} + \frac{ (d-3)\,p_{d-3}(j)}{d-1+2j}\,.
\end{align*} 
Taking the sum in $j$ from $0$ to $\ell$, 
in view of \eqref{eq:exactsummation} 
the contribution of the first term on the r.h.s. can be computed explicitely
\begin{align*}
    (d-1)! \, (d-2) \sum_{j=0}^{\ell} \frac{\mu_j}{(2j+d-1)^2} 
    &= \frac{p_{d-2}(\ell)}{2} + \frac{(d-2)(d-3)}{2} \left(\sum_{j=0}^{\ell} \frac{p_{d-3}(j)}{d-1+2j} \right)\,.
\end{align*}
To treat the second term on the r.h.s.,  we first apply 
\eqref{eq:exact recursion pm} to each summand individually
\begin{align}
\label{eq:intermediate result summation}
    (d-1)! \, (d-2) \sum_{j=0}^{\ell} \frac{\mu_j}{(2j+d-1)^2} 
    &= \frac{p_{d-2}(\ell)}{2} + \frac{d-3}{2} 
    \left(\sum_{j=0}^{\ell} \frac{p_{d-2}(j)-p_{d-2}(j-1)}{d-1+2j} \right)\,.
\end{align}
Now we use Abel's formula of summation by parts, i.e. 
\begin{align*}
\sum_{j=0}^{\ell} A_j (B_j - B_{j-1}) = A_\ell B_\ell + \sum_{j=0}^{\ell-1} (A_j - A_{j+1})B_j\,,
\quad A_j,B_j\in\R\,,\ B_{-1}=0\,,
\end{align*}
with the choice $A_j \coloneqq (d-1+2j)^{-1}$ and $B_j \coloneqq p_{d-2}(j)$. If $\ell=0$ we use the convention that sums of the type 
$\sum_{j=0}^{-1}$ vanish.
This gives
\begin{align*}
\sum_{j=0}^{\ell} \frac{p_{d-2}(j)-p_{d-2}(j-1)}{d-1+2j}=
\frac{p_{d-2}(\ell)}{d-1+2\ell}+2\sum_{j=0}^{\ell-1}
\frac{p_{d-2}(j)}{(d-1+2j)(d+1+2j)}\,.
\end{align*}
In view of the elementary estimate
\begin{align*}
4(j+1)(d-2+j)\leq (d-1+2j)(d+1+2j)
\end{align*}
and the definition of the Pochhammer symbol we claim that
\begin{align*}
\frac{p_{d-2}(j)}{(d-1+2j)(d+1+2j)}\leq  
\frac{p_{d-2}(j)}{4(j+1)(d-2+j)}=
\frac{p_{d-3}(j)}{4(j+1)}=
\frac{p_{d-4}(j+1)}{4}
\end{align*}
and by \eqref{eq:exactsummation}
\begin{align*}
\sum_{j=0}^{\ell} \frac{p_{d-2}(j)-p_{d-2}(j-1)}{d-1+2j}
&\leq 
\frac{p_{d-2}(\ell)}{d-1+2\ell}+\frac{1}{2}\sum_{j=0}^{\ell-1}
p_{d-4}(j+1)
\\
&=\frac{p_{d-2}(\ell)}{d-1+2\ell}+\frac{p_{d-3}(\ell)-p_{d-3}(0)}{2(d-3)}
\,.
\end{align*}
Finally, we make use of the identities 
$p_{d-2}(\ell)=(d-2+\ell)p_{d-3}(\ell)$ and $p_{d-3}(0)=(d-3)!$
and get
\begin{align*}
\frac{d-3}{2}
\sum_{j=0}^{\ell} \frac{p_{d-2}(j)-p_{d-2}(j-1)}{d-1+2j}
\leq 
\frac{(d-3)p_{d-2}(\ell)}{2(d-1+2\ell)}
+
\frac{p_{d-2}(\ell)}{4(d-2+\ell)}-\frac{(d-3)!}{4}
\,.
\end{align*}
Inserting this back into \eqref{eq:intermediate result summation}
we finally arrive at
\begin{align}\label{eq:finalstep2}
 (d-1)! \, (d-2) \sum_{j=0}^{\ell} \frac{\mu_j}{(2j+d-1)^2} 
\leq  \alpha\,p_{d-2}(\ell) -\frac{(d-3)!}{4}\,.   
\end{align}
with
\begin{align}\label{eq:finalstep2alpha}
\alpha\coloneqq \frac12+\frac{(d-3)}{2(d-1+2\ell)}+\frac{1}{4(d-2+\ell)}\,.
\end{align}

{\em Step 3.} We turn now to the full expression in
\eqref{eq:Riesz means formula gamma=1a}--\eqref{eq:Riesz means formula gamma=1b}. 
Using the identity 
\begin{align*}
d! N_\ell =  \frac{(d+2\ell) (d+\ell-1)!}{\ell! } =
(d+2\ell) (d+ \ell -1) p_{d-2}(\ell)
\end{align*}
together with \eqref{eq:finalstep2} we find that 
\begin{align}
    \mathrm{Tr} \left( -\Delta-\frac{\kappa}{|x|} + \Lambda \right)_- 
     &= \sum_{j = 0}^{\ell } 
    \frac{\mu_j\kappa^2}{(2j+d-1)^2} - N_\ell \Lambda \\
    &\leq  \frac{\Lambda \, \eta^d}{d! (d-2)} p_{d-2}(\ell)\, \omega - \frac{\kappa^2}{4(d-1)(d-2)^2} \,.
    \label{eq:Estimated Riesz means}
\end{align}
Here 
\begin{align*}
\omega\coloneqq  d \eta^{2-d} \alpha - \eta^{-d}(d-2) \beta  \,,
\end{align*}
where
$\alpha$ is given in \eqref{eq:finalstep2alpha} and 
$\beta \coloneqq (d+2\ell)(d+\ell-1)$.
We can now use the fact that
\begin{align*}
\omega\leq  
    \sup_{\eta \geq  0}\, (d\eta^{2-d}  \alpha - \eta^{-d} (d-2)  \beta) 
    = 2\, {\alpha^{\frac{d}{2}}}{\beta^{1-\frac{d}{2}}}
\end{align*}
to further estimate \eqref{eq:Estimated Riesz means} from above. 
We obtain    
\begin{align}
    \mathrm{Tr} \left( -\Delta-\frac{\kappa}{|x|} + \Lambda \right)_-  
    \leq  \frac{2^{2-d}\Lambda \, \eta^d}{d! (d-2)} \, G(\ell) - \frac{\kappa^2}{4(d-1)(d-2)^2} 
    \label{eq:LT esitmate final}
\end{align}
with $\ell=\lceil (\eta+1-d)/2 \rceil-1 \in \N_0$ as before and $G$ being defined as 
\begin{align}
    G(t)\coloneqq p_{d-2}(t)
    \left(1 + \frac{d-3}{d-1+2t}+ \frac{1}{2 (d-2+t)}\right)^{\frac{d}{2}}\left(\left(\frac{d}{2}+t\right)\left(d +t -1\right)\right)^{1-\frac{d}{2}} \,,\quad t\geq  0\,. \label{eq:bigGfunction}
\end{align}
The coefficient in front of $G(\ell)$ is precisely
the value of the semi-classical phase
space integral, see \eqref{eq:LT r.h.s. gamma=1}.
The upcoming Lemma \ref{lem:polynomial} shows that $G(\ell)\leq  1$ for all $\ell \in \N_0$. 
Consequently, \eqref{eq:LT strict ineq gamma1} is proven.
\end{proof}

\begin{lemma} \label{lem:polynomial} For $d \geq  4$, the function $G$ defined in~\eqref{eq:bigGfunction} 
is strictly increasing for $t\geq  0$ and satisfies $\lim_{t \ra +\infty} G(t) = 1$.
\end{lemma}

\begin{proof} To simplify some calculations, it is more convenient to work with the translated function $g(t) \coloneqq G\left(t-\frac{d-1}{2}\right)$, which after an index shift in the definition of the Pochhammer symbol is given as follows
\begin{align*}
     g(t) = \left[\prod_{k=0}^{d-3} \left( t- \frac{d-3}{2} +k \right)\right]\,\left(1 + \frac{d-3}{2t}+ \frac{1}{2 (t + \frac{d-3}{2})}\right)^{\frac{d}{2}} \left(\left(t+\frac{1}{2}\right)\left(t + \frac{d -1}{2}\right)\right)^{1-\frac{d}{2}} \,.
\end{align*}
Our goal is then to show that $g(t) \leq  1$ for $t\geq  {(d-1)}/{2}$ and $\lim_{t\ra +\infty} g(t) = 1$. 

The limit is immediate to compute.  To prove the inequality, it suffices to show that the derivative of $\log g(t)$ is non-negative for $t\geq  {(d-1)}/{2}$. To this end, we first note that
$(\log g)'(t) $ is given by
\begin{align}
\sum_{k=0}^{d-3} \left( \frac{1}{t-\frac{d-3}{2}+k} \right)  + \frac{d}{2}\left(\frac{4 t + 2d-5}{2t^2 + (2d-5)t + \frac{(d-3)^2}{2}} - \frac{1}{t} - \frac{1}{t+\frac{d-3}{2}}\right) - \frac{d-2}{2}\left(\frac{1}{t+\frac{1}{2}}+ \frac{1}{t+\frac{d-1}{2}}\right)
\,. \label{eq:loggderivative}
\end{align}
Next, we use two elementary inequalities to find a simpler lower bound for $(\log g)'(t)$. The first inequality is immediate to verify and reads as follows
\begin{align}
    \frac{4t+2d-5}{2t^2+(2d-5)t+\frac{(d-3)^2}{2}} 
 \geq  \frac{4t+2d-5}{2t^2 + (2d-5)t + \frac{(d-2)(d-3)}{2}} = \frac{1}{t+\frac{d-2}{2}} + \frac{1}{t+\frac{d-3}{2}} \quad \mbox{for $t\geq  0$.}  \label{eq:firstred}
\end{align}
The second inequality follows from the relation between 
harmonic and arithmetic means:
\begin{align}
    \sum_{k=1}^{d-3} \left(t-\frac{d-3}{2}+k \right)^{-1}
    \geq (d-3)^2\left(\sum_{k=1}^{d-3} 
    \left(t-\frac{d-3}{2}+k\right)\right)^{-1}=
    \frac{d-3}{t+\frac{1}{2}}
    \quad \mbox{for }t\geq  \frac{d-3}{2}.
    \label{eq:convexity}
\end{align}
Hence, applying \eqref{eq:firstred} and~\eqref{eq:convexity} to \eqref{eq:loggderivative}, we find that 
\begin{align*}
(\log g)'(t) &\geq 
\frac{1}{t-\frac{d-3}{2}} +  \frac{d-3}{t+\frac{1}{2}}+\frac{d}{2}
\left(\frac{1}{t+\frac{d-2}{2}}-\frac{1}{t}\right)- \frac{d-2}{2}\left(\frac{1}{t+\frac{1}{2}}+ \frac{1}{t+\frac{d-1}{2}}\right)
\\
&=\frac{1}{t-\frac{d-3}{2}}-\frac{\frac{d}{2}}{t}+ \frac{\frac{d}{2}-2}{t+\frac{1}{2}}+\frac{\frac{d}{2}}{t+\frac{d-2}{2}}-
 \frac{\frac{d}{2}-1}{t+\frac{d-1}{2}}
\\
&=\frac{h(t)}{(t-\frac{d-3}{2})t(t+\frac12)(t+\frac{d-2}{2})(t+\frac{d-1}{2})} 
\end{align*}
with 
\begin{align*}
    h(t) = \frac{1}{4}(d-2)(d-4) t^2 + \frac{1}{16}(5d-16)(d-1)(d-2)t + \frac{1}{32}d(d-1)(d-2)(d-3)\,.
\end{align*}
One can easily see that all coefficients of $h(t)$ are non-negative for $d\geq  4$. Therefore, we obtain $(\log g)'(t) \geq  h(t) \geq  0$ for $t\geq  (d-1)/2$, which completes the proof. 
\end{proof}

\section{On the sharp CLR inequalities for the shifted Coulomb Hamiltonian}\label{sec:CLR} 
\begin{proof}[Proof of Theorem \ref{thm: CLR Coulomb}]
By \eqref{eq:Nl} and \eqref{eq:LT r.h.s. gamma=0}, we need to study the behaviour of the quotient 
\begin{align}
    R_d(\eta)
    \coloneqq 
    \frac{\mathrm{Tr} \left( -\Delta - \frac{\kappa}{|x| } + \Lambda \right)_-^0 }{L_{0,d}^\mathrm{cl} \int_{\RR^d}\left(\frac{\kappa}{|x|}-\Lambda\right)_+^{d/2}\mathrm{d}x}
    = \frac{d+2 \ell }{2^{1-d} \eta^d} \prod_{j=1}^{d-1} (\ell + j)\,. \label{eq: def R_d}
\end{align}
Here we use again the notation $\eta = \kappa/\sqrt{\Lambda}>0$ and $\ell=\lceil\tau\rceil-1$
with $\tau \coloneqq (\eta+1-d)/2$. Observe that, in view of $\tau \geq  \ell$,  we have 
\begin{align}
    R_d(\eta) 
    = \frac{d+2 \ell }{2^{1-d} \eta^d}\prod_{j=1}^{d-1} ( \ell + j)
    \leq  \frac{\tau + \frac{d}{2}}{\left( \tau + \frac{d - 1}{2} \right)^d}  \prod_{j = 1}^{d - 1} (\tau + j)
    = Q_d(\tau) \label{ineq: C_d bounded by Q_d}
\end{align}
with $Q_d$ defined in \eqref{eq:Qfunction}. 
 Moreover, we note that, even though $Q_d(\tau) \neq R_d(2 \tau + d-1)$ for any $\tau >0$, we have the equality
\begin{align*}
     Q_d(\tau_0) = \lim_{\tau \downarrow \tau_0} R_d(2\tau + d-1) \quad \mbox{for any $\tau_0 \in \N_0$.}
\end{align*}

As $R_d(2 \tau + d-1)$ is strictly decreasing in any interval $(\tau_0,\tau_0+1]$ with $\tau_0 \in \N_0$, the identity above implies that the supremum of $Q_d$ over $\N_0$ gives the optimal excess factor in the CLR inequality.
For a visual illustration of $Q_d$ and $R_d$ see Figure~\ref{fig:Rd vs Qd}.
The proof is then completed with the upcoming Lemma~\ref{lemma:Property Q_d}.
\end{proof}
\begin{figure}[ht]
    \centering
    \begin{minipage}{0.49\textwidth}
        \centering
        \resizebox{1\linewidth}{!}{\input{plots/Rd_vs_Qd_d5.pgf}} 
        
    \end{minipage}
    \begin{minipage}{0.49\textwidth}
        \centering
        \resizebox{1\linewidth}{!}{\input{plots/Rd_vs_Qd_d6.pgf}}
        
    \end{minipage}
    \caption{Comparison of $R_d(2\tau+d-1), Q_d(\tau)$ for $d=5,6$. The bold marked dots emphasize where $R_d$ takes values. }
    \label{fig:Rd vs Qd}
\end{figure}
 
\begin{lemma} \label{lemma:Property Q_d} 
For $d= 3$, the function $Q_d$ defined in \eqref{eq:Qfunction} is strictly decreasing in $(-1,+\infty)$. 
For $d \geq  4$, $Q_d$ has a unique maximum at $t^\ast_d \in (-1,+\infty)$ satisfying 
\begin{align}
    \frac{d^2}{6} - \frac{3d}{2} + \frac{7}{3} < t^\ast_d < \frac{d^2}{6} -\frac{d}{2} -\frac{2}{3}\,. \label{eq:Maxbounds}
\end{align}
\end{lemma}

\begin{proof}[Proof of Lemma \ref{lemma:Property Q_d}] 
{\em Step 1.} Let us define the set  
\begin{align*}
    P \coloneqq \left\{-\tfrac{d}{2}, -\tfrac{d-1}{2} \right\} \cup \{ -(d-1),-(d-2),\dots,-1\}
    = \left\{ - \left\lceil \tfrac{d}{2} \right\rceil + \tfrac{1}{2} \right\} \cup \{ -(d-1),-(d-2),\dots,-1\}\,,
\end{align*}
which consists of $d$ elements.
Then, for $t\in \R \backslash P$ an explicit calculation yields the representation 
\begin{align}
\label{eq:Qderiv}
   Q_d'(t) = Q_d(t)\, f_d(t)  
\end{align}
with $f_d \, : \R \backslash P \ra \R$ given by
\begin{align}
    f_d (t) \coloneqq \frac{1}{t+\frac{d}{2}} - \frac{d}{t + \frac{d-1}{2}} + \sum_{k=1}^{d-1} \frac{1}{t+k}
    \label{eq:simplifiedderivative}\,.
\end{align}

For $d=3$, one can immediately verify that $f_3(t)<0$ for all $t>-1$. Since $Q_3(t)>0$ for $t>-1$, we have $Q_3'(t)<0$ for $t>-1$ and
the case $d=3$ is proven. 

Consider now $d \geq  4$.
Since $Q_d(t)>0$ for $t>-1$, by \eqref{eq:Qderiv} it suffices to prove that $f_d$ has a unique zero in $(-1,+\infty)$ located in the interval \eqref{eq:Maxbounds}, where it changes from positive to negative values for increasing argument. 
To this end, we rewrite the rational function $f_d$ as a quotient of co-prime polynomials, i.e., 
\begin{align*}
    f_d(t)=\frac{p(t)}{q(t)}\,.
\end{align*}
The set of poles of $f_d(t)$
matches the set $P$ and all these poles are of order one. Hence, 
the polynom $q(t)$ is of degree $d$ and can be chosen as follows
\begin{align*}
    q(t) \coloneqq \left(t+ \left\lceil \frac{d}{2} \right\rceil - \frac{1}{2}\right) \prod_{k=1}^{d-1} (t+k)\,.
\end{align*} 
Due to the cancellation of the terms of order $1/t$ at infinity in \eqref{eq:simplifiedderivative} we have $f_d(t)=\mathcal{O}(t^{-2})$ as $t\to\infty$, and 
the degree of the
polynomial $p$ is at most $d-2$. In fact, a calculation provided in Appendix~\ref{app:proof} shows that 
\begin{align} 
p(t) = -\frac{d}{2} \biggr(t^{d-2} + \left(\frac{d^2}{3} - \left\lfloor\frac{d}{2}\right\rfloor - \frac{1}{3}  \right) t^{d-3} + r(t)\biggr), \label{eq:coefficients}
\end{align}
where $r$ is a polynomial of degree at most $d-4$. 

{\em Step 2.} Every zero of $f_d$ must be a zero of $p$ and,
in particular, the function $f_d$ has at most $d-2$ zeros.
Analysing the sign changes of $f_d$ in the intervals between two consecutive poles, we can infer the location of the zeros of $f_d$.
Indeed, from~\eqref{eq:simplifiedderivative} we conclude that  
\begin{align}
\label{eq:normalpoles}
 \lim_{t \uparrow \tau_k} f_d(t) 
    = -\infty &
    \quad \text{and}\quad  
    \lim_{t\downarrow \tau_k} f_d(t) = + \infty 
    \quad \mbox{for}\ \tau_k\in P \setminus\{-(d-1)/2\}\,,
    \end{align}
as well as 
\begin{align*}
    \lim_{t \uparrow -\frac{d-1}{2}} f_d(t) 
    = + \infty &
    \quad\text{and}\quad
    \lim_{t \downarrow -\frac{d-1}{2}} f_d(t)
    = -\infty \,.
\end{align*}
Let us define the disjoint open intervals
$I_j=(-j-1,-j)$ for $j=1,\dots,d-1$ with $j\not\in\{\lfloor \tfrac{d}{2}\rfloor -1,\lfloor \tfrac{d}{2}\rfloor\}$ and
\begin{align*}
    I_j \coloneqq \begin{dcases} 
    \left( - \tfrac{d}{2}, - \left\lfloor \tfrac{d}{2}\right\rfloor +1 \right) \quad &\mbox{for}\ j = \lfloor \tfrac{d}{2}\rfloor -1\,,\\
    \left( -\left\lfloor \tfrac{d}{2}\right\rfloor-1,-\tfrac{d}{2}\right) \quad &\mbox{for}\ j = \lfloor \tfrac{d}{2}\rfloor\,.\end{dcases}
\end{align*}
Note that $-(d-1)/2\in I_j$ for $j={\lfloor \tfrac{d}{2}\rfloor -1}$,
while
each open interval $I_j$ for $j \in \{1,...,d-2\} \setminus \{ \lfloor d/2 \rfloor -1\}$ has two poles of the type
\eqref{eq:normalpoles} as its endpoints and does not contain any poles inside. 
By the  intermediate value theorem we see 
that $f_d(t)$ has at least one zero in each of the latter intervals.
This means that $f_d(t)$ has at least $d-3$ distinct zeros inside the interval $(-d+1,-1)$. 

Moreover, the leading coefficient of $p$ is negative and therefore
\begin{align*}
    \lim_{t \rightarrow +\infty} p(t) = -\infty\,.
\end{align*}
Since $q(t)>0$ for $t>-1$, we see that 
$f_d(t)$ is negative for sufficiently large $t$. On the other hand,
in view of $d\geq 4$ we have 
$\lim_{t \downarrow -1} f_d(t) = +\infty$. Again, by the intermediate value theorem we deduce that $f_d$ has at least one zero in the interval $(-1,+\infty)$. 
Since $f_d$ has at most $d-2$ zeros in total,  we conclude that $f_d$ must have exactly one zero in each of the intervals $I_j$ for $j \in \{1,...,d-2\} \setminus \{ \lfloor d/2\rfloor -1\}$ and additionally
one zero in the interval $(-1,+\infty)$ (see Figure~\ref{fig:f6} for an example of $f_d$), where it changes from positive to negative sign
as the argument increases. Hence,
this rightmost zero corresponds to the unique local maximum of $Q_d$
in the interval $(-1,+\infty)$.
\begin{figure}[ht]
    \centering
    \begin{minipage}{0.49\textwidth}
        \centering
        \resizebox{1\linewidth}{!}{\input{plots/f_6.pgf}} 
        
    \end{minipage}
    \caption{The function $f_d$ for $d=6$ and its four zeros: 3 negative ones and 1 positive one, as stated in the proof of Lemma \ref{lemma:Property Q_d}.}
    \label{fig:f6}
\end{figure}

{\em Step 3.} We now proceed to prove the bound in~\eqref{eq:Maxbounds}.  To this end,
note that the previous arguments do not only give us a qualitative picture of the function $f_d$, but they also allow for a quantitative estimate on the negative zeros of the polynomial $p$. 

From the previous discussion we know that $p$ has together with $f_d$
exactly $d-2$ distinct zeros, one in each interval $I_j$
for $j \neq \lceil  (d-1)/2 \rceil -1$ and one zero $t_d^\ast>-1$.
Since $p$ has the degree $d-2$, all these zeros are of order 
one. Therefore, taking the leading coefficient in \eqref{eq:coefficients} into account, we can write $p$ in the factorized form
\begin{align}
    p(t) = -\frac{d}{2} (t - t_d^\ast) \prod_{k = 1, \, k\neq \lfloor \frac{d}{2}\rfloor -1}^{d-2} (t + k + \varepsilon_k), \label{eq:p in factorised form}
\end{align}
where all $\varepsilon_k \in (0,1)$.\footnote{Note that
we include the factor for $k=\lfloor \frac{d}{2}\rfloor$
directly in the product, as all points from 
$I_{\lfloor \frac{d}{2}\rfloor}=\left(\frac{d}{2},\lfloor \frac{d}{2}\rfloor +1\right)$ permit the necessary representation as well.}

To estimate $t_d^\ast$ we compare \eqref{eq:p in factorised form} with~\eqref{eq:coefficients} and apply 
Vieta's formula applied to the term of order $t^{d-3}$.
This gives 
\begin{align*}
     - t_d^\ast + \sum_{k=1, \, k \neq \lfloor \frac{d}{2}\rfloor-1}^{d-2} (k+\varepsilon_k) = \frac{d^2}{3} - \left\lfloor \frac{d}{2} \right \rfloor - \frac{1}{3}\,.
\end{align*}
In view of $0 < \varepsilon_k < 1$, the 
upper and lower bounds~\eqref{eq:Maxbounds} follow. 
\end{proof}

\begin{remark}\label{rem:Q*}
Since by our analysis 
$Q_d(t)$ is strictly increasing for all $t>t_d^\ast$ and
$\lim_{t\to+\infty}Q_d(t)=1$, we conclude that $Q_d(\ell)>1$ 
for all $\ell\in\N$ with $\ell>t_d^\ast$. Therefore, $Q_d^\ast>1$
for all $d\geq 3$.
\end{remark}

Let us now turn to the proof of Proposition~\ref{cor:asymptotic}.

\begin{proof}[Proof of Proposition~\ref{cor:asymptotic}]
By Lemma~\ref{lemma:Property Q_d} for $d\geq 4$ 
the point $t^*_d$ at which 
$Q_d$ maximizes over the positive real numbers satisfies
\begin{align*}
\frac{d^2}{6}-\frac{3d}{2}+\frac{7}{3}\leq t^*_d \leq \frac{d^2}{6}-\frac{d}{2}-\frac{2}{3}\,.
\end{align*}
Let us estimate $\log Q_d(t^*_d)$. For this we use the Taylor expansion
of $Q_d(t)$ around $t^*_d$. Since $f_d$ is the logarithmic derivative
of $Q_d$ and $f_d(t^*_d)=0$, the Lagrange remainder term gives for $t=d^2/6$
\begin{align}\label{eq:Lagrange1}
    \left|\log Q_d\left(\frac{d^2}{6}\right) - \log Q_d(t^\ast_d)\right| 
    \leq  \frac12 \sup_{s \in [t_d^\ast,\frac{d^2}{6}]} |f'_d(s)| \cdot \left|\frac{d^2}{6}-t_d^\ast\right|^2 \leq  \frac{9d^2}{8}\sup_{s \in [t_d^\ast,\frac{d^2}{6}]} |f'_d(s)|\,.
\end{align}
By \eqref{eq:simplifiedderivative}
the derivative $f'_d(s)$ 
is equal to
\begin{align*}
f'_d(s) 
&= -\frac{1}{(s+\frac{d}{2})^2} + \frac{d}{(s+\frac{d-1}{2})^2}- \sum_{k=1}^{d-1} \frac{1}{(s+k)^2} \\
&= \frac{4s-1+2d}{4\,(s+\frac{d}{2})^2(s+\frac{d-1}{2})^2} + \frac{d-1}{(s+\frac{d-1}{2})^2} - \sum_{k=1}^{d-1} \frac{1}{(s+k)^2}\,,\quad s\in\R\setminus P\,.
\end{align*} 
Note that for $s>0$ the sum $\sum_{k=1}^{d-1} {(s+k)^{-2}}$ can be estimated from above and below as follows
\begin{align}\label{eq:upperlowerboundsum}
    \frac{d-1}{(s+1)(s+d)}\, =  \int_1^{d} \frac{1}{(s+k)^2} \mathrm{d} k\,   
    \leq \, 
    \sum_{k=1}^{d-1} \frac{1}{(s+k)^2} \,
    \leq   
    \int_0^{d-1} \frac{1}{(s+k)^2} \mathrm{d} k\, 
    = \,\frac{d-1}{t(s+d-1)}\,.
\end{align}
In particular, the lower bound in \eqref{eq:upperlowerboundsum} leads,
lets say for 
values $s$ from $\mathcal{I}=\left[\frac{d^2}{6}-\frac{3d}{2},\frac{d^2}{6}\right]$,
to 
\begin{align}
\notag
    f_d'(s) 
    & \leq  \frac{4s-1+2d}{4\,(s+\frac{d}{2})^2(s+\frac{d-1}{2})^2} + \frac{d-1}{(s+\frac{d-1}{2})^2} - \frac{d-1}{(s+1)(s+d)}\\
    & = \frac{4s-1+2d}{4\,(s+\frac{d}{2})^2(s+\frac{d-1}{2})^2} + \frac{(d-1)\left(2s + d - (\frac{d-1}{2})^2\right)}{(s+\frac{d-1}{2})^2 (s+1)(s+d)}
    \leq  C_+d^{-5}\,,\quad s\in \mathcal{I}\,,
    \label{eq:upperasymptoticboundf}
\end{align}
with some uniform constant $C_+>0$ for all sufficiently large $d$. 
In a similar way  the upper bound in \eqref{eq:upperlowerboundsum} gives for 
the same admissible values of $s$
\begin{align}
\notag
    f_d'(s) 
    & \geq  \frac{4s-1+2d}{4\,(s+\frac{d}{2})^2(s+\frac{d-1}{2})^2} + \frac{d-1}{(s+\frac{d-1}{2})^2} - \frac{d-1}{s\,(s+d-1)}\\
    & = \frac{4s-1+2d}{4\,(s+\frac{d}{2})^2(s+\frac{d-1}{2})^2} - \frac{(d-1)^3}{4\,(s+\frac{d-1}{2})^2\,s\,(s+d-1)}\geq -C_-d^{-5}
    \,,\quad s\in \mathcal{I}\,,
    \label{eq:lowerasymptoticboundf}
\end{align}
with some uniform constant $C_->0$ for all sufficiently large $d$.
Thus, by \eqref{eq:Lagrange1}, \eqref{eq:upperasymptoticboundf} and
\eqref{eq:lowerasymptoticboundf} it holds
\begin{align*}
    \left|\log Q_d\left(\frac{d^2}{6}\right) - \log Q_d(t^\ast_d)\right|=\mathcal{O}(d^{-3})
    \quad\mbox{as}\quad d\ra+\infty\,.
\end{align*}

By the way $f_d$ changes sign at $t_d^*$ we see that 
$Q_d(t)$ is strictly increasing for $0\leq  t<t_d^*$ and strictly decreasing for
$t>t_d^*$. Hence the maximum of $Q_d$ over all non-negative integers 
is taken at $\ell=\lfloor t_d^*\rfloor$ or  at 
$\ell=\lceil t_d^*\rceil$. Using again the Taylor expansion
of $Q_d(t)$ at $t_d^*$, \eqref{eq:upperasymptoticboundf} and
\eqref{eq:lowerasymptoticboundf} we see that
\begin{align}
\label{eq:asympqlt}
    \left|\log Q_d(\ell) - \log Q_d(t^\ast_d)\right| 
    \leq  \frac12 \sup_{s \in [\lfloor t_d^\ast\rfloor,\lceil t_d^\ast\rceil]}
    |f'_d(s)| \cdot \left|\ell-t_d^\ast\right|^2 \leq  C d^{-5}\quad\text{as}\quad d\ra+\infty\,. 
\end{align}
Therefore,
\begin{align*}
    Q_d(\ell)  = Q_d \left(\frac{d^2}{6}\right) (1 + \mathcal{O}(d^{-3}))
    \quad\mbox{as}\quad d\ra+\infty\,.
\end{align*}
It remains to compute that 
\begin{align*}
Q_d \left(\frac{d^2}{6}\right)=1+\frac{3}{2d}+\frac{45}{8d^2}+\mathcal{O}(d^{-3})
 \quad\mbox{as}\quad d\ra+\infty\,.
\end{align*}
This is best done computing the asymptotics for $\log Q_d(d^2/6)$ and inserting the result into the Taylor expansion for
the exponential function at zero.
\end{proof}

\section{On the CLR conjecture}\label{sec:CLR conj}

\begin{proof}[Proof of Theorem~\ref{thm:CLRconjecture}] 

Let $A_d$ be the function given in \eqref{eq:ad}.
For $t\in \R \backslash \{ -1,-2,\dots, -d+1 \}$
the derivative of $A_d$ satisfies the identity $A'_d(t) = A_d(t) g_d(t)$ with 
\begin{align*}
    g_d(t)\coloneqq \frac{1-\frac{d}{2}}{t+\frac{d}{2}} - \frac{\frac{d}{2}}{t+\frac{d}{2}-1} + \sum_{k=1}^{d-1} \frac{1}{t+k}\,. 
\end{align*}
As $A_d$ is positive for $t\geq 0$, it suffices to study the behavior of $g_d$ for $t\geq 0$. For $d \leq 4$, it is immediate to see that $g_d \leq 0$ and therefore $A_d$ is decreasing, which completes the proof in this case. 

For $d \geq 5$, we divide the proof into two cases. First, we consider the case when $d$ is even. Later, we deal with the case when $d$ is odd. In the even case, the proof follows the exact same steps of the proof of Theorem~\ref{thm: CLR Coulomb}. For convenience of the reader, we sketch these steps below.

{\em Step 1: the case $d> 5$ even.} 
For $d$ even, we note that the poles of the first two terms also appear as poles within the sum term. 
Rewriting $g_d$ as a quotient of co-prime polynomials, we obtian 
\begin{align*}
    g_d(t) = \frac{p(t)}{q(t)}, 
\end{align*}
where  $ q(t) \coloneqq \prod_{k=1}^{d-1} (t+k) $ and $p(t)$ is a polynomial of order $d-3$  satisfying 
\begin{align}
p(t) = -\frac{d}{2} \left(t^{d-3} + \left(\frac{d^2}{3} -d+ \frac{2}{3}\right) t^{d-4} + \mathcal{O}(t^{d-5})\right). \label{eq:polynomialform}
\end{align}
Thus, $p$ (and consequently $g_d$) has at most $d-3$ zeros. 
Analysing the behavior of $g_d$ near its poles and using the intermediate value theorem, as we did in the proof of Theorem~\ref{thm: CLR Coulomb}, we conclude that $g_d$ has at least one zero in each of the intervals $(-k-1,-k)$ for $k\in \{1,...,d-2\}\setminus \{ d/2, d/2-2\}$, and at least one zero in the interval $(-1, +\infty)$. 
Therefore, $g_d$ (and consequently $p$) has exactly one zero of order one 
in each of the described intervals. 
Hence, $p(t)$ can be written as a product as follows
\begin{align}
    p(t) = -\frac{d}{2} (t-\tilde{t}_d)\prod_{ k=1, \, k \notin  \left\{\frac{d}{2},\frac{d}{2}-2 \right\} }^{d-2} (t+k+\varepsilon_k) , \label{eq:factorised_p_conjecture}
\end{align}
where $\tilde{t}_d$ is the unique zero in the interval $(-1,+\infty)$. 
Expanding the r.h.s. of \eqref{eq:factorised_p_conjecture} then yields 
\begin{align}
   p(t) = -\frac{d}{2} t^{d-3} + \frac{d}{2} \left(\tilde{t}_d - \sum_{ k=1, \, k \notin  \left\{\frac{d}{2},\frac{d}{2}-2 \right\} }^{d-2}  (k+\varepsilon_k)\right) t^{d-4} + \mathcal{O}(t^{d-5}) . \label{eq:fac_p_conj_expand}
\end{align}
By comparing the coefficient 
in front of $t^{d-4}$
in \eqref{eq:fac_p_conj_expand} with the corresponding 
one in~\eqref{eq:polynomialform} and using the estimates $0\leq  \varepsilon_k \leq  1$, we conclude that
\begin{align*}
   \frac{d^2}{6} - \frac{3d}{2} +\frac{7}{3} \leq \tilde{t}_d \leq \frac{d^2}{6} -\frac{d}{2} -\frac{5}{3},
\end{align*}
which is within the interval in Theorem~\ref{thm:CLRconjecture}.

{\em Step 2: the case $d\geq 5$ odd.} 
For $d$ odd, we can not apply the same argument because $g_d(t)$ has five consecutive poles with different sign, namely $\{-\frac{d+1}{2},-\frac{d}{2},-\frac{d-1}{2},-\frac{d}{2}+1,-\frac{d-3}{2}\}$, and therefore the counting poles argument does not add up. Hence, we shall first use a simple estimate to get rid of the middle pole $t = \{-\frac{d-1}{2}\}$, and then proceed as in the previous step. To this end, it is more convenient for calculations to work with the shifted function
\begin{align} \label{eq:def tilde gd}
	\tilde{g}_d(s) 
    \coloneqq 
    g_d\left(s-\frac{d-1}{2}\right) 
    = 
    \frac{1-\frac{d}{2}}{s+\frac{1}{2}} - \frac{\frac{d}{2}}{s-\frac{1}{2}} + \sum_{j= -\frac{d-3}{2}}^{j=\frac{d-1}{2}} \frac{1}{s+j} \, .
\end{align}
Note that it suffices to study the behaviour of the shifted function for $s\geq \frac{d-3}{2}$. Moreover, the set of poles of $\tilde{g}_d(s)$ is $P \coloneqq \{-\frac{d-1}{2}, -\frac{d-1}{2}+1,...,-1,-\frac12, 0, \frac12, 1, ..., \frac{d-3}{2}\}$, and the pole at $t = -\frac{d-1}{2}$ is now located at $s=0$. The simple estimate we use to get rid of this pole is the following
\begin{align}
    \frac{1-a_d}{s-\frac12} + \frac{a_d}{s+\frac12}  \leq \frac{1}{s} \leq \frac{1/2}{s-\frac12} + \frac{1/2}{s+\frac12}, \quad \mbox{for $s\geq \frac{d-3}{2}$,} \label{eq: funny est}
\end{align}
where $a_d \coloneqq \frac12 +\frac{1}{2(d-3)}$. 
Precisely, we note that by~\eqref{eq: funny est} we have
\begin{align} \label{eq:squeeze}
    h_{a_d}(s) 
    < \tilde{g}_d(s) <
    h_{1/2}(s) 
    \quad \text{for} \quad 
    s> \frac{d-3}{2},
\end{align}
where $h_a(s)$ is the function defined by
\begin{align*}
    h_a(s) 
    \coloneqq 
    \frac{1-\frac{d}{2}}{s+\frac12} - \frac{\frac{d}{2}}{s-\frac12}  + \frac{1-a}{s-\frac12} + \frac{a}{s+\frac12} + \sum_{\substack{k=-\frac{d-3}{2}, \,  k \neq 0}}^{\frac{d-1}{2}} \frac{1}{s+k}
    \quad \text{for }
    s \in (\mathbb{R} \backslash P)\cup \{0\}. 
\end{align*} 
We can now apply the "counting poles" argument to the function $h_a(s)$ for $0 \leq a\leq 1$ to show that $h_a(s)$ has a unique zero inside the interval $(\frac{d-3}{2},+\infty)$ and to localize this zero. More precisely, by following the arguments in the previous step, one can show (see Appendix~\ref{app:more calculations}) that 
\begin{align}
    h_a(s) \geq 0 \quad \mbox{for} \quad s\leq \left(\frac{d-1}{2} + a\right)^{-1} \left( \frac{d^3-6d^2+11d-3}{12} - \frac{d-2}{2} a\right)\label{eq:lowerbound}
\end{align}
and
\begin{align}
    h_a(s) \leq 0 \quad \mbox{for} \quad s \geq \left(\frac{d-1}{2} + a\right)^{-1} \left( \frac{d^3-6d^2+11d-3}{12} - \frac{d-2}{2} a\right)+ d-3 . \label{eq:upperbound}
\end{align}
Hence, by setting~ $a = 1/2$ in~\eqref{eq:upperbound} and using~\eqref{eq:squeeze}, we find that
\begin{align*}
    \tilde{g}_d(s) < 0 \quad \mbox{for}\quad s \geq \frac{d^2}{6} - \frac53 + \frac{1}{2d}.
\end{align*}
Shifting back, $g_d(t) = \tilde{g}_d(t+\frac{d-1}{2})$, and using the trivial estimate $\frac{1}{2d} \leq \frac16$ we obtain
\begin{align}
    g_d(t) < 0 \quad \mbox{for}\quad t \geq \frac{d^2}{6} - \frac{d}{2}-1. \label{eq:upperbound2}
\end{align}
Similarly, by setting $a= a_d = \frac12 + \frac{1}{2(d-3)}$ in~\eqref{eq:lowerbound} and using~\eqref{eq:squeeze} we find
\begin{align*}
    \tilde{g}_d(s) > 0 \quad \mbox{for}\quad \frac{d-3}{2}< s \leq \frac{d^2}{6} - d + \frac76 + \frac{3d-10}{6(d^2-3d+1)}.
\end{align*}
Shifting back and using the trivial estimate $\frac{3d-10}{6(d^2-3d+1)} \geq 0$ (valid for $d\geq 4$) we get 
\begin{align}
    g_d(t) > 0 \quad \mbox{for}\quad -1 < t \leq  \frac{d^2}{6} - \frac{3d}{2} +\frac{5}{3}. \label{eq:lowerbound2}
\end{align}
In particular, any zeros of $g_d(t)$ for $t>-1$ are within the interval in Theorem~\ref{thm:CLRconjecture}. Together with~\eqref{eq:upperbound2} and~\eqref{eq:lowerbound2}, this implies that the maximum of $A_d(t)$ is achieved inside the desired interval.

{\em Step 3: proving the asymptotic expression \eqref{eq:CLRconjecture_expansion}.}
Let now $t\geq 0$. 
We note that
\begin{align*}
    \frac{A_d(t)}{Q_d(t)} = 
    \frac{(t+\frac{d-1}{2})^d}{(t+\frac{d}{2})^{\frac{d}{2}} 
    (t+ \frac{d}{2}-1)^{\frac{d}{2}}}
    =
    \left(1+\frac{1}{\left(2t+d-1\right)^2-1}\right)^{\frac{d}{2}}\,.
\end{align*}
Thus, one has $A_d(t)>Q_d(t)$. 
We know the maximizing points of both $Q_d$ and $A_d$ are contained 
within the interval $[d^2/6-3d/2,d^2/6]$. For such $t$
we see that 
\begin{align*}
    \log A_d(t) - \log Q_d(t) 
    = \frac{d}{2}\log  \left(1+\frac{1}{\left(2t+d-1\right)^2-1}\right)
    \leq  C d^{-3}\,,
    \quad t\in \left[\frac{d^2}{6}-\frac{3d}{2},\frac{d^2}{6}\right]\,,
\end{align*}
with some uniform $C>0$ for sufficiently large $d$. This implies the first equality in \eqref{eq:CLRconjecture_expansion}. 
The second equality in \eqref{eq:CLRconjecture_expansion} follows from the expansion from Proposition~\ref{cor:asymptotic}.
\end{proof}

\appendix
\section{Proof of \texorpdfstring{\eqref{eq:coefficients}}{(1)}}
\label{app:proof}

\begin{proof}[Proof of equation~\eqref{eq:coefficients}]
First let us rewrite $f_d$ from \eqref{eq:simplifiedderivative} as follows
\begin{align*}
f_d(t)&=\frac{1}{t+\frac{d}{2}} - \frac{1}{t+\frac{d-1}{2}} + \sum_{k=1}^{d-1} \left(\frac{1}{t+k} - \frac{1}{t+ \frac{d-1}{2}}\right)
=-\frac{1}{2\left(t+\frac{d-1}{2}\right)\left(t+\frac{d}{2}\right)}+\sum_{k=1}^{d-1}
\frac{\frac{d-1}{2}-k}{\left(t+\frac{d-1}{2}\right)(t+k)}\,.
\end{align*}
Multiplying this expression by $q(t)$ yields
\begin{align}
    p(t) &= f_d(t)\,\left(t+ \left\lceil \frac{d}{2} \right\rceil - \frac{1}{2}\right)\, \prod_{k=1}^{d-1} (t+k)
    \\
    &= -\frac{1}{2} \prod_{k=1, \, k\neq \lfloor \frac{d}{2}\rfloor}^{d-1} (t+k) + \sum_{k = 1 }^{d-1} \left(\frac{d-1}{2} - k\right) \frac{t+\frac{d}{2}}{t+\lfloor\frac{d}{2}\rfloor}\prod_{j = 1, \, j \neq k}^{d-1} (t+j) \,.
     \nonumber
\end{align}
Note that the two different shapes the summand 
for $k=\lfloor\frac{d}{2}\rfloor$ takes for odd and even 
$d$ can be combined into one expression
\begin{align*}
\left(\frac{d-1}{2} - \left\lfloor \frac{d}{2} \right\rfloor\right)
\prod_{k=1, \, k\neq \lfloor \frac{d}{2}\rfloor}^{d-1} (t+k)\,.
\end{align*}
Hence, we have $p(t)=I(t) + J(t)$ with
\begin{align*}
    I(t) &\coloneqq \left(\frac{d-1}{2} - \left\lfloor \frac{d}{2} \right\rfloor -\frac{1}{2} \right)\prod_{k=1, \, k\neq \lfloor \frac{d}{2}\rfloor}^{d-1} (t+k), \\
    J(t) &\coloneqq \sum_{k=1, \, k \neq \lfloor \frac{d}{2} \rfloor}^{d-1}\left(\frac{d-1}{2} - k\right)  \left(t+\frac{d}{2}\right) \prod_{j=1, \, j \notin \{ k,  \lfloor \frac{d}{2}\rfloor \} }^{d-1} (t+j).
\end{align*}

The term $I(t)$ expands as follows
\begin{align*}
    I(t) &= \left(\frac{d}{2} - \left\lfloor \frac{d}{2} \right\rfloor -1 \right)  t^{d-2} + T_1(d)  t^{d-3} + r_1(t)\,,
\end{align*} 
where the polynomial $r_1$ is at most of degree $d-4$. 
Let
\begin{align*}   
    S(d) \coloneqq \sum_{k=1,\, k \neq \lfloor \frac{d}{2}\rfloor}^{d-1} k=
    \frac{(d-1)d}{2}-\left\lfloor\frac{d}{2}\right\rfloor\,.
\end{align*}
By Vieta's theorem the factor in front of $t^{d-3}$ in $I(t)$ is then given by
\begin{align*}
T_1(d)=\left(\frac{d}{2} - \left\lfloor \frac{d}{2} \right\rfloor -1 \right)\,S(d)=
\left\lfloor\frac{d}{2}\right\rfloor^2+
\left(-\frac{d^2}{2}+1\right)\left\lfloor\frac{d}{2}\right\rfloor
+\frac{d^3}{4}-\frac{3d^2}{4}+\frac{d}{2}\,.
\end{align*}

Regarding $J(t)$, we find that 
\begin{align*}
    J(t) =& \sum_{k=1, \, k \neq \lfloor \frac{d}{2} \rfloor}^{d-1} \left(\frac{d-1}{2} - k\right)\left( t^{d-2} + \left(\frac{d}{2} + \sum_{j=1, \, j \notin \{ k,  \lfloor \frac{d}{2}\rfloor \} }^{d-1} j \right) t^{d-3} + r_{2,k}(t)\right)\\
    =& \left(1-d+\left\lfloor\frac{d}{2}\right\rfloor\right) t^{d-2}  + T_2(d)\, t^{d-3} + r_3(t)\,,
\end{align*}
where $r_{2,k}$ and $r_3$ are polynomials of degree of at most $d-4$.
We compute the factor in front of $t^{d-3}$ in $J(t)$ as follows
\begin{align*}
T_2(d)\,=&
\sum_{k=1, \, k \neq \lfloor \frac{d}{2} \rfloor}^{d-1} \left(\frac{d-1}{2} - k\right)\,\left(\frac{d}{2} + S(d) - k \right)
\\
=\,&\frac{(d-2)(d-1)}{2}\left(\frac{d}{2}+S(d)\right)-
\left(d-\frac{1}{2}+S(d)\right)S(d)+ \frac{(d-1)d(2d-1)}{6}
-\left\lfloor\frac{d}{2}\right\rfloor^2
\\
=\,&-2\left\lfloor\frac{d}{2}\right\rfloor^2+
\left(\frac{d^2}{2}+\frac{3d}{2}-\frac{3}{2}\right)\left\lfloor\frac{d}{2}\right\rfloor
-\frac{5d^3}{12}+\frac{d^2}{2}-\frac{d}{12}\,.
\end{align*}

Combining these results we find that
\begin{align*}
I(t)+J(t)=-\frac{d}{2}t^{d-2}+(T_1(d)+T_2(d))t^{d-3}+(r_1(t)+r_3(t))\,,
\end{align*}
where
\begin{align*}
T_1(d)+T_2(d)=-\left\lfloor\frac{d}{2}\right\rfloor^2
+\left(\frac{3d}{2}-\frac{1}{2}\right)\left\lfloor\frac{d}{2}\right\rfloor
-\frac{d^3}{6}-\frac{d^2}{4}+\frac{5d}{12}
=-\frac{d}{2}\left(\frac{d^2}{3}-\left\lfloor\frac{d}{2}\right\rfloor -\frac{1}{3}\right)\,.
\end{align*}
This implies~\eqref{eq:coefficients}.
\end{proof}

\section{Additional calculations} \label{app:more calculations}

\subsection{Asymptotics of \texorpdfstring{$Q_d$}{(1)} }

We first compute the asymptotics of $\log Q_d(t)$. To this end, we write
\begin{align*}
	\log Q_d(t) &= \log\left(\frac{t+\frac{d}{2}}{t+\frac{d-1}{2}}\right) + \sum_{j=1}^{d-1} \log\left(\frac{t+j}{t+\frac{d-1}{2}}\right)\\
	&= \log\left(1+\frac{1}{2t + d-1}\right) + \sum_{j=1}^{d-1} \log\left(1+\frac{2j - (d-1)}{2t+d-1}\right). 
\end{align*}
If $t\geq d^2/6-\frac{3d}{2}$, the fraction inside the first logarithmic term is of order $d^{-2}$, while the fractions in the logarithms inside the sum can be of order $d^{-1}$. As the sum has $d-1$ terms, to capture all terms of order up to $d^{-2}$, it suffices to make a linear expansion of the first logarithmic term and a third order expansion of logarithms inside the sum. Precisely, we use the expansion
\begin{align*}
\log(1+x) = 1 + x - \frac{x^2}{2} + \frac{x^3}{3} + \mathcal{O}(x^4) \quad \mbox{as $x\rightarrow 0$,}
\end{align*}
to get
\begin{align}
\log Q_d(t) = \frac{1}{2t+d-1} + \frac{1}{2t+d-1} \sum_{j=1}^{d-1} (2j -d +1) - \frac{1}{2(2t+d-1)^2} \sum_{j=1}^{d-1} (2j-d+1)^2  \label{eq:sum} \\
 + \frac{1}{3(2t+d-1)^3} \sum_{j=1}^{d-1} (2j-d+1)^3 + \mathcal{O}(d^{-3}). \nonumber
\end{align}
Since the summand in the last sum is monotonically increasing, we can compare this sum against the integrals
\begin{align*}
	\int_1^{d-1} (2j-d+1)^3 \mathrm{d} j = \frac{(2j-d+1)^4}{8}\biggr\rvert_{j=1}^{d-1} = \mathcal{O}(d^3),\\ 
    \quad 
    \int_0^{d} (2j-d+1)^3 \mathrm{d} j = \frac{(2j-d+1)^4}{8}\bigr\rvert_{j=0}^{d-2} = \mathcal{O}(d^3), 
\end{align*}
to conclude that, for $t \geq d^2/6 - \frac{3d}{2}$, the total contribution of the last term in~\eqref{eq:sum} is of order $\mathcal{O}(d^{-3})$ and therefore unimportant for us. 

We now evaluate the remaining sums explicitly. For the first sum, a simple calculation yields
\begin{align}
\sum_{j=1}^{d-1}  (2j - d +1) = d(d-1) - (d-1)^2 = d-1.  \label{eq:sum1}
\end{align}
For the second sum, we have
\begin{align}
\sum_{j=1}^{d-1} (2j-d+1)^2 &= \sum_{j=1}^{d-1} \left(4j^2 - 4 j (d-1) + (d-1)^2 \right)= \sum_{j=1}^{d-1} \left(4j(j+1) - 4j d + (d-1)^2\right) \nonumber \\
&= 4 \sum_{j=1}^{d-1}\frac{j(j+1)(j+2) - (j-1)j (j+1)}{3} - 4d \sum_{j=1}^{d-1} j  + (d-1)^3 \nonumber \\
&= 4 \frac{j(j+1)(j+2)}{3}\biggr\rvert_{j=0}^{d-1} - 2 d^2(d-1) + (d-1)^3 \nonumber \\
&=\frac{4}{3} d(d^2-1) - (d-1)(d^2+2d-1) = \frac{d^2}{3} - d^2 + \mathcal{O}(d). \label{eq:sum2}
\end{align}
Since we are free to choose $t=\frac{d^2}{6} + r_d$ for any $r_d$ of order $d$, it is convenient to take $t_d\coloneqq \frac{d^2}{6} - \frac{d-1}{2}$ to simplify the denominator in~\eqref{eq:sum}. Hence, with this choice and by plugging the expressions~\eqref{eq:sum1} and~\eqref{eq:sum2} into~\eqref{eq:sum}, we obtain
\begin{align*}
	\log Q_d(t_d) &= \frac{1}{d^2/3} + \frac{d-1}{d^2/3} - \frac{\frac{d^3}{3} - d^2}{2 \frac{d^4}{9}} + \mathcal{O}(d^{-3}) =\frac{3}{2d} +\frac{9}{2 d^2} + \mathcal{O}(d^{-3}).
\end{align*}
In particular $\log Q_d(t_d) \leq C d^{-1}$ for $C>0$ large enough. Therefore, we can use the second order Taylor  expansion of the exponential function at $x=0$, namely
\begin{align*}
\exp(x) = 1 + x + \frac{x^2}{2} + \mathcal{O}(x^3),
\end{align*}
to conclude that
\begin{align*}
Q_d(t^\ast_d) &= Q_d(t_d) + \mathcal{O}(d^{-3}) = \exp(\log Q_d(t_d)) + \mathcal{O}(d^{-3}) = 1 + \log Q_d(t_d) + \frac{\log Q_d(t_d)^2}{2} + \mathcal{O}(d^{-3})\\
&= 1 + \frac{3}{2d} + \frac{9}{2d^2} + \frac{9}{8 d^2} +\mathcal{O}(d^{-3}) = 1 + \frac{3}{2d} + \frac{45}{8 d^2} + \mathcal{O}(d^{-3}).
\end{align*}

\subsection{Calculations on CLR conjecture for $d$ odd}

In this section we prove formulas~\eqref{eq:lowerbound} and~\eqref{eq:upperbound}. First, by the definition of $h_a$ we have
\begin{align*}
    h_a(s) 
    &  = \left(1-\frac{d}{2}\right) \frac{2s}{s^2-\frac14} - \frac{a}{s^2-\frac14} + \frac{1}{s+\frac{d-1}{2}} + \sum_{\substack{k = - \frac{d-3}{2},\,  k \neq 0}}^{\frac{d-3}{2}} \frac{1}{s+k} \\
    &= \left(1-\frac{d}{2}\right) \frac{2s}{s^2-\frac14} - \frac{a}{s^2-\frac14} + \frac{1}{s+\frac{d-1}{2}}  + \sum_{k=1}^{\frac{d-3}{2}} \frac{2s}{s^2-k^2} 
    \, . 
\end{align*}
Then we rewrite $h_a(s)$ as a quotient of co-prime polynomials, i.e. 
\begin{align*}
    h_a(s) = \frac{p_a(s)}{q(s)} 
    \quad \mbox{where} \quad 
    q(s) = \left(s^2-\frac14\right) \left(s+\frac{d-1}{2}\right) \prod_{k=1}^{\frac{d-3}{2}} (s^2-k^2) 
\end{align*}
and
\begin{align}
 \label{eq:p expansion}
\begin{aligned}
    p_a(s) = &-\left((d-2)s +a\right) \left(s+\frac{d-1}{2}\right) \prod_{j=1}^{\frac{d-3}{2}} (s^2-j^2) + \left(s^2-\frac14\right) \prod_{j=1}^{\frac{d-3}{2}} (s^2-j^2) \\
    & + 2s\left(s^2-\frac14\right) \left(s+\frac{d-1}{2}\right) \sum_{j=1}^{\frac{d-3}{2}} \prod_{\substack{k=1,  \,  k \neq j}}^{\frac{d-3}{2}} (s^2-j^2).
\end{aligned}
\end{align}
To compute the coefficients of the leading terms of $p_a(t)$, we denote $C(d) \coloneqq (d-1)(d-2)(d-3)/24$ and observe that 
\begin{align*}
    \prod_{j=1}^{\frac{d-3}{2}}(s^2-j^2) 
    &= 
    s^{d-3} - s^{d-5} \sum_{j=1}^{\frac{d-3}{2}} j^2 + \mathcal{O}(s^{d-7}) 
    = 
    s^{d-3} - {C(d)} s^{d-5} + \mathcal{O}(s^{d-7}) \, , \\
  \prod_{\substack{k=1, \, k \neq j}}^{\frac{d-3}{2}} (s^2-k^2) &= \frac{1}{s^2-j^2} \left(s^{d-3} + C(d) s^{d-5} +  \mathcal{O}(s^{d-7})  \right)\\
    &= \left(s^{-2} + j^2 s^{-4} + \mathcal{O}(s^{-6})\right)\left( s^{d-3}  - C(d) s^{d-5}+ \mathcal{O}(s^{d-7})\right)\\
    &= s^{d-5} +\left(j^2-C(d)\right) s^{d-7} + \mathcal{O}(s^{d-9}) \, \quad \mbox{as $s \rightarrow \infty$.}
\end{align*}
Plugging these two identities into~\eqref{eq:p expansion} we find
\begin{align}
    p_a(s) =& -\left((d-2)s+a\right)\left(s+\frac{d-1}{2}\right) \left(s^{d-3} - C(d) s^{d-5}\right) + \left(s^2-\frac14\right) \left(s^{d-3} - C(d) s^{d-5}\right)\nonumber \\
    &+ \left(s^3-\frac{s}{4}\right)\left(s+\frac{d-1}{2}\right) \sum_{j=1}^{\frac{d-3}{2}} 2\left(s^{d-5} - C(d) s^{d-7} + j^2 s^{d-7}\right) +r_a(s) \nonumber \\
    =& -\left(\frac{d-1}{2} +a \right) s^{d-2} + \left(2 C(d) - \frac{d-1}{2} a - \frac{d-2}{4}\right) s^{d-3} +r_a(s) \nonumber \\
    =& - \left(\frac{d-1}{2} + a\right) s^{d-2} + \left(\frac{d^3-6d^2 + 8d}{12} - \frac{d-1}{2} a\right)s^{d-3} + r_a(s) \, , \label{eq:formula p_a}
\end{align}
where $r_a(s)$ is a polynomial of order at most $d-4$.

We can now inspect the poles of $h_a(s)$ and use the same arguments from the previous proof to conclude that, for $d\geq 7$, $h_a(s)$ has exactly one zero in each of the following intervals
\begin{align*}
	I_j = 
    \begin{dcases} \left(j-1-\tfrac{d-1}{2},j-\tfrac{d-1}{2}\right) \quad &\mbox{for $1 \leq j \leq \frac{d-3}{2}$,}\\
	\left(-\tfrac{1}{2},\tfrac{1}{2} \right) \quad &\mbox{for $j=\frac{d-1}{2}$,}\\
	\left(j-\tfrac{d-1}{2}, j +1- \tfrac{d-1}{2}\right) \quad &\mbox{for $\frac{d+1}{2} \leq j \leq d-3$.} \end{dcases}
\end{align*}
For $d = 5$, the same holds but the last interval should be excluded since $\frac{d+1}{2} > d-3$ in this case. In any case, note that $h_a(s)$ and $p_a(s)$ share the same zeros. 
Since $p_a(s)$ is a polynomial of order $d-2$ satisfying $\lim_{s\ra +\infty} p_a(s) = - \infty$ and $\lim_{s \downarrow \frac{d-3}{2}} p_a(s) = +\infty$, it must have one last zero $\tilde{s}(a)$ inside the interval $\big( \frac{d-3}{2}, +\infty \big)$. 
Therefore, $p_a(s)$ can be written as
\begin{align}
    p_a(s) &= -\left(\frac{d-1}{2} + a\right) (s-\tilde{s}(a))\left(s+\frac12+\varepsilon_0\right)\prod_{\substack{k=-\frac{d-1}{2}, \, k \not \in \{-1,0\}}}^{\frac{d-5}{2}} (s-k-\varepsilon_k) \nonumber \\ 
    &= -\left(\frac{d-1}{2} +a \right) \left(s^{d-2} - \left(\tilde{s}(a) - \frac{1}{2} + \varepsilon_0 + \sum_{\substack{k=-\frac{d-1}{2}, \, k \not \in \{-1,0\}}}^{\frac{d-5}{2}} k+\varepsilon_k \right) s^{d-3}\right) + r_a(t) \nonumber \\ 
    &= -\left(\frac{d-1}{2} + a\right) \left(s^{d-2} + \left((d-3)-\delta -\tilde{s}(a) + \frac12\right) s^{d-3}\right) + r_a(s), \label{eq:p_a for estimation}
\end{align}
where $\varepsilon_0, \varepsilon_k \in (0,1)$ and $0 \leq \delta \leq d-3$. 
Comparing coefficients of the term $s^{d-3}$ in~\eqref{eq:formula p_a} and in~\eqref{eq:p_a for estimation}, we find that
\begin{align*}
    \tilde{s}(a) & = (d-3-\delta) + \frac12 + \left(\frac{d-1}{2} +a\right)^{-1}\left(\frac{d^3-6d^2+8d}{12} - \frac{d-1}{2} a\right) \\
    & =  (d-3-\delta) + \left(\frac{d-1}{2} + a\right)^{-1} \left( \frac{d^3-6d^2+11d-3}{12} - \frac{d-2}{2} a\right) \, .
\end{align*}
Therefore, $h_a(s)$ has a unique zero for $s \geq \frac{d-3}{2}$ and this zero is inside the interval of length $d-3$ with the left endpoint at
\begin{align*}
    \left(\frac{d-1}{2}+a\right)^{-1} \left(\frac{d^3-6d^2+11d -3}{12} - \frac{d-2}{2} a\right). 
\end{align*}
As $\lim_{s \downarrow \frac{d-3}{2}} p_a(s) = +\infty$, $\lim_{s \ra \infty} p_a(s) = -\infty$ and $q(s) \geq 0$ for $s\geq (d-3)/2$, this implies that~\eqref{eq:upperbound} and ~\eqref{eq:lowerbound} hold. 
\addtocontents{toc}{\protect\setcounter{tocdepth}{-1}}
\section*{Acknowledgements}
The authors thank Professor Heinz Siedentop for valuable discussions and 
for calling attention to the relation between the second coefficient in Proposition~\ref{remark:improvedestimate} and the Scott correction term in $d=3$, which lead us to Proposition~\ref{thm:LTnextorder}.

T.C.~Corso acknowledges funding by the \emph{Deutsche Forschungsgemeinschaft} (DFG, German Research Foundation) - Project number 442047500 through the Collaborative Research Center "Sparsity and Singular Structures" (SFB 1481). 

\addtocontents{toc}{\protect\setcounter{tocdepth}{2}}




\bigskip

\begin{thebibliography}{}


\bibitem[AL78]{AL78}
 	\textsc{M. Aizenman} and \textsc{E.H. Lieb},
    \newblock On semi-classical bounds for eigenvalues of Schrödinger operators. 
    \newblock \doi{10.1016/0375-9601(78)90385-7}
    {\emph{Physics Letters A}} \textbf{66}, (1978), no.6, 427--429.
    \newblock \mr{0598768}.
    \hfill 



  \bibitem[CCR24]{CCR24}
 	\textsc{T.C.~Corso} and \textsc{T. Ried},
 	\newblock {\emph{On a variational problem related to the Cwikel-Lieb-Rozenblum and Lieb-Thirring inequalities}}.
 	\newblock \doi{10.1007/s00220-024-05216-y}{\emph{Communications in Mathematical Physics}} \textbf{406} (2025), no. 50.
    \newblock \mr{4861103}.
    \newblock \zbl{07986041}.
 	\hfill

\bibitem[Cor24]{Cor24}
 	\textsc{T.C.~Corso},
 	\newblock {\emph{A generalized three lines lemma in Hardy-like spaces}}.
 	\newblock \href{https://arxiv.org/abs/2407.10117}{arxiv 2407.10117}
 	\hfill

\bibitem[Cwi77]{Cwi77}
 	\textsc{M.~Cwikel},
    \newblock Weak Type Estimates for Singular Values and the Number of Bound States of Schrodinger Operators. 
    \newblock \doi{10.2307/1971160}
    {\emph{Annals of Mathematics}} \textbf{106}, (1977), no.1, 93--100.
    \newblock \mr{0473576}.
    \newblock \zbl{0362.47006}.
    \hfill 

  



\bibitem[FLW23]{FLW23}
	\textsc{R.L.~Frank, A.~Laptev}, and \textsc{T.~Weidl},
	\newblock \doi{10.1017/9781009218436}{\emph{Schr\"odinger operators: eigenvalues and Lieb-Thirring inequalities}}.
	\newblock Cambridge Studies in Advanced Mathematics Vol.\ 200, Cambridge University Press, Cambridge, 2023.
	\newblock \mr{4496335}.
	\newblock \zbl{07595814}.
	\hfill

 
\bibitem[Fra21]{Fra21}
 	\textsc{R.L.~Frank},
 	\newblock \doi{10.1090/pspum/104/01877}{The Lieb--Thirring inequalities: recent results and open problems}.
 	\newblock In \emph{Nine mathematical challenges—an elucidation} (eds.\ A.~Kechris, N.~Makarov, D.~Ramakrishnan, X.~Zhu),
 	\newblock  Proceedings of Symposia in Pure Mathematics \textbf{104}, American Mathematical Society, Providence, RI, 2021, pp. 210--235.
 	\newblock \mr{4337417}.
	\newblock \zbl{1518.35002}.
 	\hfill
 

\bibitem[GGM78]{GGM78}
 	\textsc{V.~Glaser, H.~Grosse,} and \textsc{A.~Martin},
 	\newblock
     Bounds on the number of eigenvalues of the Schr\"odinger operator.
     \newblock \doi{10.1007/BF01614249}{\emph{Communications in Mathematical Physics}} \textbf{59}  (1978), 197--212
 	\newblock \mr{491613}.
	\newblock \zbl{0373.35050}.
 	\hfill
  



	


 







\bibitem[Lie80]{Lie80} 
	\textsc{E.H.~Lieb},
   \newblock The number of bound states of one-body Schroedinger operators and the Weyl problem. 
	\newblock In \emph{Geometry of the Laplace operator}, Proceedings of Symposia in Pure Mathematics Vol.\ 36, American Mathematical Society, Providence, RI, 1980, 241--252.
   \newblock \mr{573436}, 
   \newblock \zbl{0445.58029}
	\hfill

\bibitem[LT76]{LT76} 
	\textsc{E.H.\ Lieb} and \textsc{W.~Thirring}, 
	\newblock Bounds on the eigenvalues of the Laplace and Schr\"odinger operators. 
	\newblock \doi{10.1090/S0002-9904-1976-14149-3}{\emph{Studies in Mathematical Physics: Essays in Honor of Valentine Bargmann (Princeton Series in Physics)}} (1976), 269-303. 
	\newblock \zbl{0342.35044}.
	\hfill 


	

\bibitem[Roz76]{Roz76}
 	\textsc{G.V.~Rozenblum},
    \newblock Distribution of the discrete spectrum of singular differential operators (Russian).
    \newblock \doi{10.2307/1971160}
    {\emph{Izv. Vysš. Učebn. Zaved. Matematika}}, (1976), no.1, 75-86.
    \newblock \mr{0430557}.
    \newblock \zbl{0249.35069}.
    \hfill 
 
 
 
 

	


\bibitem[Sel24]{Sel24}
    \textsc{A.D.~Selvi},
    \newblock Cwikel–Lieb–Rozenblum inequalities for the Coulomb Hamiltonian. 
    \newblock \doi{10.1007/s13324-024-00870-w}
    {\emph{Analysis and Mathematical Physics}} \textbf{14}, (2024).
    \newblock \mr{4701844}.
    \newblock \zbl{1535.35046}.



\bibitem[Sol96]{Sol96}
 	\textsc{A.V.~Sobolev},
    \newblock Discrete spectrum asymptotics for the Schrödinger operator in a moderate magnetic field. 
    {\emph{Reviews in Mathematical Physics}} \textbf{8}, (1996), no.6, 861--903.
    \newblock \mr{1365349}.
    \newblock \zbl{0843.35093}.
    \hfill 


 


\bibitem[SW87]{SW87}
 	\textsc{H.~Siedentop} and \textsc{R.~Weikard},
    \newblock On theLeading Energy Correction for the Statistical Model of the Atom: Interacting Case. 
    \newblock \doi{10.1007/BF01218487}
    {\emph{Communications in Mathematical Physics }} \textbf{112}, (1987), no.3, 471--491.
    \newblock \mr{0908549}.
    \newblock \zbl{0920.35120}.
    \hfill 


\bibitem[Wei96]{Wei96} 
	\textsc{T. Weidl}, 
	\newblock On the Lieb-Thirring constants $L{\gamma,1}$ for $\gamma \geqslant 1/2$.
	\newblock \doi{10.1007/BF02104912}{\emph{Communications in Mathematical Physics}} \textbf{178} No.1 (1996), 135--146. 
	\newblock \mr{2714989}.
	\newblock \zbl{0858.34075}.
	\hfill 





\end{thebibliography}
\end{document}